\documentclass[apj]{emulateapj}
\usepackage{apjfonts}
\slugcomment{Accepted by the Astronomical Journal}

\shorttitle{H$\alpha$-Selected Galaxies at
$(0.05<z<0.15)$}
\shortauthors{Kranz et al.}

\begin{document}

\title{The Homogeneous Properties of H$\alpha$-Selected Galaxies at
$(0.05<z<0.15)$}

\author{Willy D. Kranz\altaffilmark{1,A}}
\author{Kim-Vy H. Tran\altaffilmark{1,2}}
\author{Lea Giordano\altaffilmark{1}}
\author{Am\'{e}lie Saintonge\altaffilmark{1}}

\altaffiltext{1}{Institute for Theoretical Physics, University of
  Z\"urich, Z\"urich, CH-8057, Switzerland}

\altaffiltext{2}{George P. and Cynthia W. Mitchell Institute for
Fundamental Physics and Astronomy, Department of Physics and
Astronomy, Texas A\&M University, College Station, TX 77843}

\altaffiltext{A}{will@physik.uzh.ch}

\begin{abstract}

We show that the H$\alpha$ line (6563\AA) alone is an extremely
effective criterion for identifying galaxies that are uniform in color
(red), luminosity-weighted age (old), and morphology
(bulge-dominated).  By combining the {\it Sloan Digital Sky Survey}
(Data Release 6) with the {\it New York University Value-Added Galaxy
Catalog}, we have photometric and spectroscopic indices for over
180,000 galaxies at $(0.05<z<0.15)$.  We separate the galaxies into
three samples: 1) galaxies with H$\alpha$ equivalent width $<0$\AA \
($i.e.$ no emission); 2) galaxies with morphological S\'ersic index
$n>2$ (bulge-dominated); and 3) galaxies with $n>2$ that are also red
in $(g'-r')$.  We find that the H$\alpha$-selected galaxies
consistently have the smallest color scatter: for example, at
$z\sim0.05$ the intrinsic scatter in apparent $(g'-r')$ for the
H$\alpha$ sample is only $0.0287\pm0.0007$ compared to $0.0682\pm0.0014$
for the S\'ersic sample.  Applying a color-cut to the $n>2$ sample
does decrease the color scatter to $0.0313\pm0.0007$, but there remains
a measurable fraction of star-forming and/or AGN galaxies (up to
9.3\%).  All of the EW(H$\alpha$)$<0$\AA\ galaxies have $n>2$, $i.e.$
they are bulge-dominated systems.  The spectra for the three samples
confirm that the H$\alpha$-selected galaxies have the highest D4000
values and are, on average, nearly twice as old as the
S\'ersic-selected samples.  With the advent of multi-object
near-infrared spectrographs, H$\alpha$ alone can be used to reliably
isolate truly quiescent galaxies dominated by evolved stellar
populations at any epoch from $z\sim0$ up to $z\sim2$.

\end{abstract}


\section{Introduction}
\label{sec.Introduction}

Early-type galaxies are observed to follow well-defined scaling
relations across a wide range in luminosity and mass, $e.g.$ they span
a narrow range in (red) color across four magnitudes of luminosity in
the local universe \citep{Baum1959.ColorMagnitudeRelation,
Visvanathan.Sandage1977.ColorMagnitudeRelation,
Sandage.Visvanathan1978.ColorMagnitudeRelation,
  Bower.et.al1992.ColorMagnitudeRelation}.  Because the shallow slope
  in their color-magnitude (CM) relation is due primarily to changes
  in metallicity rather than age
  \citep{Faber1973.RedSequence,Kodama1997.CMrelation,Gallazzi.et.al2006.ageMetalGalsLocalUniverseSDSS,
  Graves.et.al.2007.RedSequence,Graves.et.al.2009.RedSequence}, their
  tight CM relation means that early-type galaxies at $z\sim0$ are a
  uniformly old population.  However, we have yet to fully understand
  how these deceptively simple galaxies assembled.

How the red (quiescent) sequence, as typically defined by early-type
galaxies, is populated places strong constraints on galaxy formation
models
\citep{Kodama1997.CMrelation,Bell.et.al2004.Combo17,Bower.et.al2006.GalaxyFormation,deLucia.et.al2007ColorMagnitudeRelationGalaxyClusters,Font.et.al2008.ColorSatelliteGals}.
For example, the number density of red sequence galaxies, $i.e.$ their
luminosity function, and how it evolves is critical for understanding
how quickly gas cools to form stars in halos as a function of halo
mass \citep{Rees1977.CoolingGasSFR}. In the observationally supported
picture of ``down-sizing''
\citep{Cowie1996.GalaxyEvolution,Tran2003.PostStarburtsGalsOnRedsequence,Kodama2004.GalaxyDownSizing,ThomasDaniel2005.EarlyTypGalFormation} where star
formation history is driven by mass, the faint end of the red sequence
should become less populated and weaker at higher redshift, yet
studies conflict as to whether the number of faint red galaxies in
clusters decreases with increasing redshift
\citep{deLucia.et.al2007ColorMagnitudeRelationGalaxyClusters,Crawford2009.RedSequenceLuminosityFunction}.
\citet{Ellis.et.al1997.RedSequence} and \citet{Mei2009.EvolutionCMR}
also find in their cluster sample that the CM relation as defined by
E/S0 members does not change up to $z \sim$ 1.

The observed properties of these quiescent galaxies are also important
for improving stellar population models, $e.g.$
\citet{Bruzual.Charlot.2003.BC03}, because they also tend to  be the
oldest galaxies.  Their ages must be consistent with
the age of the universe at a given redshift, and their ages are
determined primarily from observed colors and spectral indices.
However, the stellar synthesis models used today to reproduce these
observed properties have significant uncertainties
\citep{Charlot1996.UncertaintiesModelingOldStellarPopulations}. Thus
in order to use the oldest galaxies to measure, $e.g.$ how the Hubble
constant evolves with redshift
\citep{Stern.et.al2009.H(z)Measurements}, we need to better calibrate
the stellar synthesis models such that the colors and spectral indices
from models match the observed properties of the oldest galaxies at
any redshift, particularly at $0<z<2$.

Part of the problem is how to identify a galaxy population with
uniformly old stellar ages at any epoch.  Most studies use morphology,
$e.g.$ visually classified elliptical and S0 galaxies, or S\'ersic
index $n$ as a proxy for morphology
(\citealt{Hogg.et.al.2004.redSequenceWithOverdensity,Bell2004.ImageRedSequenceGalsDustyOrOld,McIntosh2005.EvolutionEarlyTypeRedSequenceGals,Cassata2005.SersicMorphology,Mei2009.CMrelationAtZ=1}).
However, there is a population of blue, star-forming galaxies that are
bulge-dominated
\citep{Menanteau2001.EvolvingSpheroidals,Cassata2005.SersicMorphology,Lintott2008.MorphologiesGalaxyZoo,Schawinski.et.al.2009GalaxyZooRedSequence},
thus early-type galaxies cannot be considered to be exclusively
passive, old systems.  Also, morphologically classifying galaxies at
higher redshifts is challenging, $i.e.$ an imaging resolution of
$<0.1''$ is required to reliably identify E/S0 galaxies
\citep{Postman2005.MorphologyAtZ=1} or measure S\'{e}rsic indices at
$z >$ 0.3 \citep{LaBarbera2002.StructuralPropertiesOfClusterGals}.  In
particular, measuring S\'ersic indices for fainter objects is
difficult because fitting surface brightness profiles is highly
dependent on the image's signal-to-noise ratio.  Alternatively,
studies have used galaxy color to separate the red sequence from the
active galaxies in the blue cloud
\citep{Baldry.et.al2004.BimodalColor-MagnitudeDistribution,Baldry2004.ColorBimodalityEvolutionImplications},
but dust can artificially redden a star-forming galaxy such that it
falls on the red sequence, $e.g.$
\cite{Wolf2005.RedGalsWithYoungStars}, and most post-starburst
galaxies lie on or near the red sequence even though they often have
H$\alpha$ emission
(\citealt{Quintero.et.al.2004.HaSelectedGalaxySample,Tran2004.E+AgalsAtIntermediateRedshifts}).

One promising solution for identifying the truly quiescent galaxies
that define the red sequence is to use spectral indices. In
particular, H$\alpha$ has been shown to be a reliable optical tracer
of a galaxy's current star formation
\citep{Kennicutt.Kent.1983.SFRnormalGals, Kennicutt.1998.HalphaSFR}
and thus can be used to identify and exclude active galaxies;
H$\alpha$ has the added benefit of not being dependent on image
resolution, although the assumption is that H$\alpha$ is integrated
over the entire galaxy.  Both \citet{Yan.et.al.2006.OIIinRedSequence}
and \citet{Graves.et.al.2009.RedSequence} use H$\alpha$ and
[OII]$\lambda3727$ to select quiescent galaxies at $0.04<z<0.1$ from
the Sloan Digital Sky Survey \citep{Stoughton.et.al.2002.SDSS}, and
they confirmed that these galaxies define a narrow red sequence.  At
higher redshift, \citet{Tran2007.SpectroscopicRedSequence} also showed
that quiescent galaxies (selected using [OII]) in MS~1054--03, a
massive galaxy cluster at $z=0.83$, define as narrow a range in color
as the elliptical galaxies identified with HST/ACS imaging
\citep{Postman2005.MorphologyAtZ=1}.  We note that there is no
physical reason to assume that absorption-line galaxies should have
the same tight color distribution, i.e. uniformity in average stellar
age, as ellipticals.

Motivated by these encouraging results, we test here H$\alpha$'s
effectiveness at identifying a uniformly aged population of galaxies
across a range of luminosity (mass).  While [OII]$\lambda3727$ is used
in most surveys to identify active galaxies at $z<1.5$, [OII] is known
to be an unreliable tracer of ongoing star formation
\citep[$e.g.$][]{Moustakas2006.SFRindicators}.  While more robust than
[OII], H$\alpha$ has yet to be widely used to identify passive
galaxies because it shifts to near-infrared wavelengths at $z>0.4$;
however, with the recent development near-infrared multi-object
spectrographs, H$\alpha$ can now be measured in galaxies up to
$z\sim2$.  The goal of our study is to lay the groundwork for using
the H$\alpha$ criterion to identify the oldest galaxies across a wide
range in redshift, $i.e.$ at $z>0.4$, and thus enable us to better
trace how the red sequence is populated as a function of redshift.

We mine the Sloan Digital Sky Survey Data Release 6 (SDSS DR6;
\citealt{Stoughton.et.al.2002.SDSS},
\citealt{Adelman-McCarthy.et.al.2008.SDSSdr6}) and the NYU Value-Added
Galaxy Catalog (NYU VAGC; \citealt{Blanton.et.al.2005.NYUcatalogue})
to select galaxies with H$\alpha$ equivalent widths of $<$
0\AA\ (absorption); we improve on the earlier work of
\citet{Yan.et.al.2006.OIIinRedSequence} and
\citet{Graves.et.al.2009.RedSequence} by extending the redshift range
to consider galaxies at $(0.05<z<0.15)$.  However, unlike
\citet{Graves.et.al.2009.RedSequence} who use a targeted combination
of spectral and morphological criteria to isolate red sequence
galaxies, we use either only spectral or only photometric criteria to
define our galaxy samples.  We compare the H$\alpha$-selected sample
to galaxies selected by S\'{e}rsic index $n$ (morphology) alone as
well as to galaxies selected by S\'{e}rsic $n$ and a color cut.  By
examining the CM relations, spectral indices, and derived ages of the
three galaxy samples, we quantify which has the highest purity where
purity is defined as the exclusion of active (star-forming) galaxies.

In \S\ref{sec.Data}, we present the data and define our selection
criteria for the Main and Luminous Galaxy samples.  The results for
the different galaxy samples are in \S\ref{section.Results}, and their
relative effectiveness is compared in \S\ref{sec.Discussion}.  Our
conclusions are in \S\ref{sec.Conclusion}.  We use a concordance
cosmology with $\Omega_{0}$ = 0.3, $\Omega_{\Lambda}$ = 0.7, $H_{0}$ =
100 $h$ km s$^{-1}$ Mpc$^{-1}$, $h$ = 1 throughout this work.

\section{Data}
\label{sec.Data}

We use spectroscopic data from the Sloan Digital Sky Survey (SDSS),
Data Release 6
(DR6)\footnote{\url{http://www.sdss.org/dr6/index.html}}
(\citealt{Stoughton.et.al.2002.SDSS};
\citealt{Adelman-McCarthy.et.al.2008.SDSSdr6}) and photometric data
from the NYU Value-Added Galaxy Catalog (NYU
VAGC)\footnote{\url{http://sdss.physics.nyu.edu/vagc/}}. The main
galaxy sample of the SDSS DR 6 includes more than 790,000 galaxy
spectra and provides one with a variety of measured spectral indices,
$e.g.$ the D4000-value (flux ratio of the 4000\AA{}-break) and the
equivalent widths (EWs) of H$\alpha$, H$\beta$ and [OIII]
(\citealt{Adelman-McCarthy.et.al.2008.SDSSdr6}). The NYU VAGC
supplements the SDSS measurements with additional data that include
the galaxy's extinction corrected Petrosian magnitude, S\'{e}rsic
index $n$, and the K-corrected absolute magnitude. The combined use of
both catalogs enables us to explore the diversity of these parameters
for a statistically large number of galaxies.

\subsection{SDSS}
\label{subsec.SDSS}

\subsubsection{DR6 Galaxy Catalog}
\label{subsubsec.DR6 catalog}

The Sloan Digital Sky Survey (SDSS) is an optical survey using the 2.5
m telescope at the Apache Point Observatory in New Mexico. SDSS uses 5
filter bands for its photometry: $u'g'r'i'z'$
(\citealt{Fukugita.et.al.1996.SDSSfilters,
Stoughton.et.al.2002.SDSS}). The SDSS also includes spectroscopic
observations with a wavelength range of $\lambda$ = 3800\AA{} -
9200\AA{} at a resolution of $\sim$ 2.7\AA{}~pixel$^{-1}$. From the
SDSS, we use the measured equivalent widths (EW) of H$\alpha$,
H$\beta$, [NII] $\lambda$6585 and [OIII] $\lambda$5007; negative EWs
denote absorption lines, and the EWs were measured after continuum
fitting.  We also use the strength of the 4000\AA-break
\citep{Bruzual.1983.D4000value} as measured by the ratio of the flux
in the 4050\AA-4250\AA\ (red) bandpass to the flux in the
3750\AA-3950\AA\ (blue) bandpass from the SDSS catalog \footnote{Note
that the SDSS catalog lists (1/D4000) values, and that the SDSS values
are measured used broader bandpasses than, e.g.
\cite{Kauffmann.et.al.2003.Dn4000value} and
\cite{GonzalezDelgado.et.al.2005.Dn4000A} who use $D_n4000$.}

\subsubsection{NYU Value-Added Galaxy Catalog}
\label{subsubsec.NYU Cataloge}

The NYU Value-Added Galaxy Catalog (NYU VAGC) is based on the SDSS DR6
catalog and provides additional photometric parameters for galaxies to
$r'$ = 18 (\citealt{Blanton.et.al.2005.NYUcatalogue}).  The
completeness limit is slightly fainter than the $r'=17.7$ stated for
DR6 \citep{Strauss.et.al.2002.SDSSmainSpecGalsSample} because the NYU
VAGC corrects for extinction using the dust maps of
\cite{Schlegel.et.el.1998.dustMaps}.  It includes S\'{e}rsic index $n$
(see \citealt{Sersic.1968.sersicIndex}) and absolute magnitude $M$ in
all five filter bands ($u'g'r'i'z'$) for nearly all SDSS DR6
galaxies. The K-corrected absolute magnitude $M$ in each filter band
has been computed via

\begin{equation} M_{f} = m_{f} - 5
\textnormal{ log}_{10}\left[D_{L}/10\textnormal{pc}\right] - K_{f}(z)
\textnormal{,} \end{equation} 

\noindent where $D_{L}$ is the luminosity distance, $f$ the filter
index and $m$ the apparent magnitude of filter $f$. In the catalog, the K-corrections $K_{f}(z)$ were calculated
for each galaxy individually with Blanton's \verb|kcorrect| software
version v4\_1\_4 (\citealt{Blanton.Roweis.2007.Kcorrection};
\citealt{Hogg.et.al.2002.Kcorrection}).

In this work we will use the NYU VAGC's apparent Petrosian magnitudes
in the $g'$ and $r'$ bands, its S\'{e}rsic index $n$ (as measured in
$r'$ imaging), and the absolute magnitude $M_{r'}$.

\subsection{Main Galaxy Sample}
\label{subsec.Main Galaxy Sample}

Our Main Galaxy sample (MG) includes only galaxies with 0.05 $< z
\leq$ 0.15. The lower redshift limit of $z$ = 0.05 is necessary to
minimize aperture problems due to the 3$^{\prime\prime}$ diameter size
of the spectroscopic fibers: at $z<0.05$ the SDSS spectra tend to only
include light from the central regions of the galaxies. The upper
redshift cutoff of $z$ = 0.15 provides a magnitude range of at least
1.5 magnitudes in a complete sample even at the higher redshift end.

To ensure completeness, we use an $r'$-band magnitude limit of $r' <$
17.7 \citep{Strauss.et.al.2002.SDSSmainSpecGalsSample}; this is
brighter than the magnitude limit in SDSS DR6 ($r'<17.77$) and the NYU
VAGC \citep[$r'<18$;][]{Blanton.et.al.2005.NYUcatalogue}.  Our
magnitude limit provides a reliably complete galaxy sample from the
combined catalogs.

The galaxies in our  Main Galaxy sample are required to have:
\begin{enumerate}
 \item a signal-to-noise ratio $>$ 3 in the $g'$- and $r'$-band photometry 
 \item a median (S/N) $>$ 5 over the entire spectrum 
 \item a measured (S/N) $>$ 2 in the H$\alpha$ line 
\end{enumerate}

\noindent Furthermore, each galaxy must have an absolute Petrosian
$r'$-band magnitude $M_{r'}$ brighter than the
K-corrected completeness limit of

\begin{equation}
M_{r',limit} = 17.7 - 5 \textnormal{ log}_{10} \left[D_{L}/10\textnormal{pc}\right] - K_{ave}(z)
\label{eq.Completeness Cut}
\end{equation}

\noindent where $D_{L}$ is the luminosity distance, 17.7 is our
applied $r'$ magnitude limit, and $K_{ave}(z)$ is the average
K-correction determined using:

\begin{equation}
K_{ave}(z) = 1.2031z - 0.0102
\label{eq.Kave}
\end{equation}

\noindent The average K-correction shown in Eq. \ref{eq.Kave} is based
on the $r'$-band K-corrections of H$\alpha$ absorption-line galaxies in
the NYU VAGC.  Earlier work by
\cite{Quintero.et.al.2004.HaSelectedGalaxySample} showed that galaxies
with H$\alpha$ absorption have an early-type spectral energy
distribution (SED) and thus there is little scatter in their $r'$-band
K-correction with redshift in the redshift range of our study (0.05 $
< z \leq$ 0.15). A $\sigma$-clipped linear least squares fit to the
H$\alpha$ absorption-line galaxies gives Eq.~\ref{eq.Kave} with a
3$\sigma$ deviation of less than 0.02 mag; note that this relation is
valid only for passive galaxies which are the focus of our study.  

The Main Galaxy sample is split into 20 redshift bins of width
$\vartriangle$\textit{z} = 0.005. The effects of passive galaxy
evolution and assuming an early-type SED to determine K-corrections on
colors and magnitudes are negligible within such narrow redshift
bins. Because we compare results of the three different selection
criteria within each redshift bin, we also minimize any bias
introduced by the $3''$ fiber aperture.  Each redshift bin's limiting
absolute Petrosian magnitude is set at the upper $z$ value. Table
\ref{tbl.Main galaxy sample} lists the 20 redshift bins with their
ranges, limiting absolute magnitude, and the number of galaxies that
meet all our selection criteria.

\begin{figure}
\epsscale{1.0}
\plotone{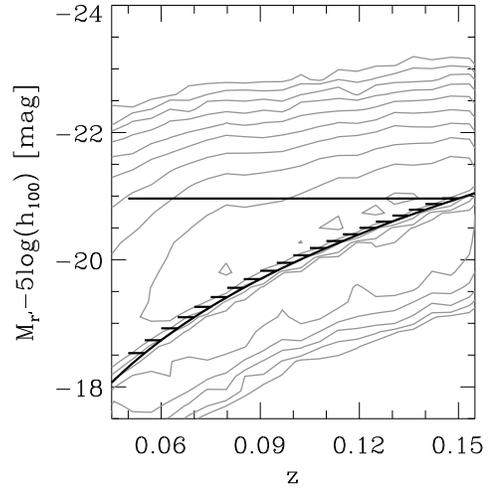}
\caption{The absolute Petrosian magnitude distribution of all NYU
Catalog galaxies from the DR6 in redshift space (gray contours). The 
solid curve is the completeness limit of the SDSS instrument in the
$r'$-band. The redshift dependent cuts in absolute magnitude
$M_{r'}$ in each redshift bin of the Main Galaxy (MG) sample
are represented by the short solid horizontal lines of width $\vartriangle$z =
0.005. The wider solid line with 
$\vartriangle$z = 0.1 is the adopted $M_{r'}$ limit of -21.0 for the Luminous
Galaxy (LG) sample (see Tables \ref{tbl.Main galaxy sample} \&
\ref{tbl.Luminous galaxy sample}). The median redshift of the SDSS
main galaxy sample is $\sim$ 0.1.}  
\label{fig.Absolute magnitudes cut-off}
\end{figure}

\subsubsection{MG1: Selecting with H$\alpha$ }
\label{subsubsec.Ha criterion}

The H$\alpha$ spectral line is the most reliable tracer of ongoing
star formation at optical wavelengths
(\citealt{Kennicutt.1998.HalphaSFR};
\citealt{Yan.et.al.2006.OIIinRedSequence}): star-forming galaxies have
significant H$\alpha$ emission, and H$\alpha$ is not strongly affected
by, $e.g.$ dust.  However, there is no a priori reason why galaxies
that do not have H$\alpha$ emission should have similar star formation
histories, i.e. have uniformly high D4000 indices.  We use H$\alpha$
to separate starforming and non-starforming galaxies and define our
first Main Galaxy sample (MG1) to have galaxies with EW(H$\alpha$) $<$
0\AA{} (Table \ref{tbl.Criteria}).  Depending on the redshift bin, the
MG1 sample can have up to $\sim5,500$ galaxies (Table
\ref{tbl.CMrelation slopes Ha<0}).

Previous studies have used both H$\alpha$ and [OII] to identify
quiescent galaxies
(\citealt{Yan.et.al.2006.OIIinRedSequence,Graves.et.al.2009.RedSequence})
and shown that this population forms a tight color-magnitude relation
(CMR). However, [OII] can be strongly dust-extincted and is known to
be an unreliable tracer of star formation
(\citealt{Jansen.et.al2001.[OII]TracerOfCurrentStarFormation,
Moustakas2006.SFRindicators}).  Also, measurements of [OII] in
galaxies at $z<0.06$ with ground-based observatories is problematic
due to absorption at $\lambda<4000$\AA~by the Earth's atmosphere.  Our
goal is to determine whether H$\alpha$ alone can identify a uniformly
quiescent galaxy population.

\subsubsection{MG2 \& MG3: Selecting with S\'{e}rsic index \& Color}
\label{subsubsec.Sersic and color}

The S\'{e}rsic index $n$ is used as a quantitative measure of a
galaxy's morphology and is determined by fitting the axisymmetric
surface brightness profile $I(r)=A \textnormal{
exp}\left[-(r/r_{0})^{1/n}\right]$ to a galaxy's surface brightness
profile (\citealt{Sersic.1968.sersicIndex}).  Galaxies with high
S\'ersic indices ($n >$ 2) tend to be bulge-dominated (early-type)
systems and those with lower indices ($n <$ 2) to be disk-dominated
(late-type) systems.  However, while S\'ersic index is a useful
measure of galaxy morphology
\citep{Sersic.1968.sersicIndex,Cassata2005.SersicMorphology,Driver2006.SersicMorphology,Wel2008.SersicMorphology},
morphology alone is not a reliable tracer of ongoing star formation
\citep[$e.g.$][]{Schawinski.et.al.2009GalaxyZooRedSequence}.  Also,
the S\'{e}rsic index $n$ depends strongly on the filter because a
galaxy's surface brightness profile varies with observed wavelength.

Multiple studies use morphology and in particular S\'{e}rsic $n>2$ to
identify red sequence galaxies, $e.g.$
\cite{Hogg.et.al.2004.redSequenceWithOverdensity}. However, this does
not provide a pure sample of red sequence galaxies: while the sample
features a strong CMR, it also includes a large fraction of blue
galaxies.  For comparison to our H$\alpha$-selected (MG1) sample, we
follow \cite{Hogg.et.al.2004.redSequenceWithOverdensity} and select
galaxies using S\'{e}rsic $n>2$ for our second Main Galaxy sample
(MG2; see Table \ref{tbl.Criteria}).  We use the S\'{e}rsic index
measured in the $r'$-band from the NYU VAGC because $r'$ is the most
sensitive of the SDSS filters.  Due to small errors in the measurement
of the local sky level, there is a bias of $\sim(-0.5)$ in the
computed S\'{e}rsic index such that a S\'{e}rsic index of $n \sim 3.5$
is measured for a real de Vaucouleurs surface brightness profile
(\citealt{Blanton.et.al.2005.NYUcatalogue},
\citealt{Blanton.et.al.2005.SDSSsersicIndex}).  Depending on redshift
bin, the MG2 sample includes up to $\sim12,000$ galaxies (Table
\ref{tbl.CMrelation slopes Sersic}), $i.e.$ about twice the number of
galaxies as in MG1.

For a more refined comparison sample, we apply an additional color
selection based on the scatter in the CMR of the H$\alpha$-selected
(MG1) sample: our third Main Galaxy (MG3) sample thus includes both a
selection on morphology (S\'{e}rsic $n>2$) and color (Table
\ref{tbl.Criteria}).  We use only galaxies with color
$\Delta(g'-r')>(-3\sigma_{OBS})$, where $\sigma_{OBS}$ is the median
absolute deviation (MAD) of the observed color scatter; because the
color distribution is asymmetric, the MAD is more appropriate for
measuring the color scatter. A more extensive discussion of the color
selection is in \S\ref{subsec.Color-magnitude relation}.  The MG3
sample has up to $\sim8,000$ galaxies in a given redshift bin (Table
\ref{tbl.CMrelation slopes Sersic+color-cut}).

\subsection{Luminous Galaxy Sample}
\label{subsubsec.Bright M Sample}

In our analysis, we test the results from the Main Galaxy sample by
comparing them to a Luminous Galaxy (LG) sample where we include all
the galaxies at 0.05 $ < z \leq$ 0.15 and apply an absolute Petrosian
magnitude limit of $M_{r'}$ = -21.0 (81,323 galaxies; see
Fig.~\ref{fig.Absolute magnitudes cut-off}).  Because the LG sample
spans a wide redshift range, we use the K-corrected rest-frame
magnitudes and colors from the NYU VAGC.  Applying the same selection
criteria as in the Main Galaxy sample, we have LG1 (EW H$\alpha<0$),
LG2 (S\'ersic $n>2$), and LG3 ($n>2$ and color-cut).  Table
\ref{tbl.Luminous galaxy sample} lists the criteria for these three
samples and how many galaxies are in each.

\subsection{Color-Magnitude Relation \& Color Scatter}\label{scatter}

To determine the color-magnitude (CM) relation in our Main and
Luminous Galaxy samples, we fit a simple linear least-squares to the
$r'$ magnitude and $(g'-r')$ color from the NYU VAGC; the fit uses the
associated photometric errors.  So that we are not weighted by
outliers, e.g. the blue cloud, we iteratively remove $3\sigma$
outliers (convergence in typically $<5$ iterations) to determine the
best fit to the red sequence.  We use the median absolute deviation
(MAD) to characterize the color scatter about the CM relation.  In the
Main Galaxy (MG) samples, we use observed magnitude and color because
we do not wish to introduce additional measurement error by converting
to rest-frame values.  However, because the Luminous Galaxy (LG)
sample spans a redshift range ($0.05<z<0.15$), we use the rest-frame
Petrosian values from the NYU VAGC.

The errors on the slope and intercept of the CM relation as well as on
the observed color scatter are determined by bootstrapping 5000
datasets from the selected galaxy sample in a given redshift bin
(\citealt{Efron.1979.Bootstrap}).  For each bootstrapped dataset, we
measure the CM relation as outlined above.  The distribution of the CM
parameters from the bootstrapped datasets then give us the $1\sigma$
errors on the slope, intercept, and the observed color scatter about
the CM relation.

Because the observed color scatter is a combination of statistical
(intrinsic) and systematic (measurement) errors, we also find it
useful to determine the intrinsic scatter for the different galaxy
samples. To determine the intrinsic scatter in color, we do the
following:

\begin{enumerate}

\item For each galaxy in the sample with magnitude $r'$, we calculate
the predicted $(g'-r')$ using the CM relation.

\item Deviations ($\delta$) in magnitude and color for the modeled CM
values are added where the random deviation is drawn from the
associated photometric errors in the NYU VAGC, i.e. $-\Delta r'\leq
\delta r'\leq\Delta r'$ where $\Delta r'$ is that galaxy's error in
magnitude.

\item A new CM relation and color scatter are determined for the
modeled galaxy sample; the original associated photometric errors for
each galaxy are assumed in fitting the new CM relation.


\end{enumerate}

We repeat this 5000 times for each galaxy sample, and the median
value of the distribution for the modeled color scatter is subtracted
in quadratures from the observed color scatter to obtain the intrinsic
scatter.  The error on the intrinsic scatter is determined by
subtracting in quadratures the $1\sigma$ of the distribution for the
modeled color scatter from the observed error.

\begin{figure}
\epsscale{1.0}
\plotone{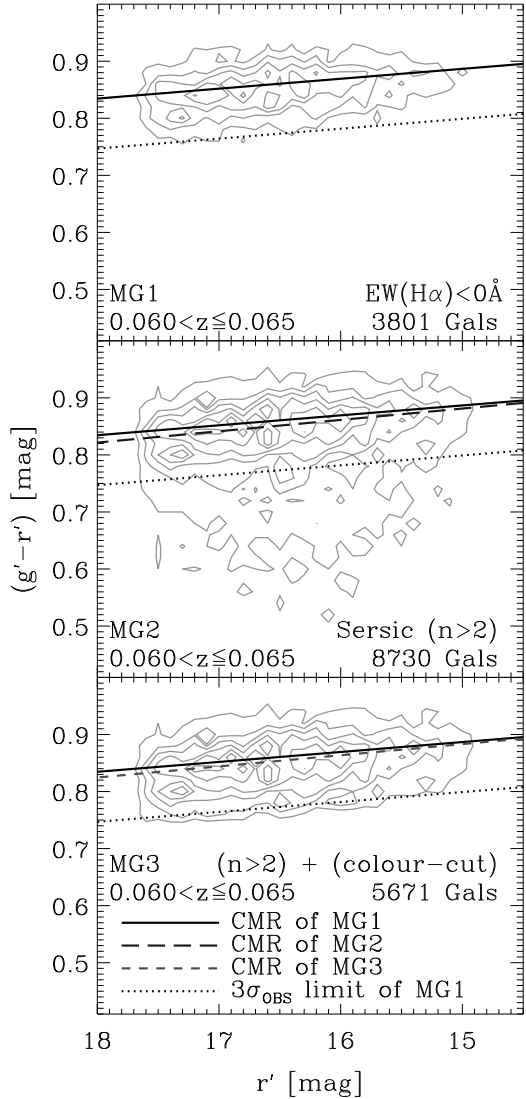}
\caption{The color-magnitude (CM) diagrams of MG1, MG2, and MG3 at
0.060 $<$ $z$ $\leq$ 0.065. All of the CM relations are sigma-clipped
least squares fits to the plotted galaxy populations. The solid line
is MG1's (upper panel) CM relation (shown in all panels), the
long-dashed line the CM relation of MG2 (middle panel), and the
short-dashed line the CM relation of MG3 (bottom panel). The dotted
line (shown in all panels) is the 3$\sigma_{\tiny{\textnormal{OBS}}}$
(MAD) limit of MG1's CM relation. The number of galaxies in each
subsample is indicated in the different panels.}
\label{fig.CMrelation slopes Ha+Sersic}
\end{figure}

\section{Results}
\label{section.Results}

\subsection{Main Galaxy Sample}

In this section, we separate our Main Galaxy sample ($0.05<z\leq0.15$)
into galaxies with H$\alpha$ absorption (MG1), S\'ersic index $n>2$
(MG2), or S\'ersic index $n>2$ and red colors (MG3), and compare their
physical properties.  Our results are listed for the 20 redshift bins
in Tables \ref{tbl.CMrelation slopes Ha<0} to \ref{tbl.averageD4000
values}.  However, for clarity, we focus on the galaxies in the
redshift bin 0.060 $< z \leq$ 0.065 in Figures \ref{fig.CMrelation
slopes Ha+Sersic} to \ref{fig.D4000.vs.Sersic Sample I}; this redshift
bin has a large number of galaxies covering a magnitude range of about
3.5 magnitudes (see Table \ref{tbl.Main galaxy sample} for the MG
sample).

\subsubsection{MG: Color-magnitude relation}
\label{subsec.Color-magnitude relation}

To judge which of our Main Galaxy samples has the most homogeneously
red galaxy population, we compare their distribution on the
color-magnitude (CM) diagram.  Galaxies with uniform ages will have a
narrow range in color while those with recent or ongoing star
formation will populate a wide range in color
(\citealt{Degiola-Eastwood1980.R-VcolorsEgals,
Merluzzi2002.CMRandAge}).  To ensure that we do not add to the
measurement error by, e.g. converting to rest-frame magnitudes by
applying k-corrections, we use apparent colors and magnitudes.

Figure \ref{fig.CMrelation slopes Ha+Sersic} (top panel) shows the
color distribution of the MG1 sample ($0.060<z<0.065$).  We determine
the CM relation with an iterative
$\sigma_{\tiny{\textnormal{MAD}}}$-clipping linear least squares fit
(see \S\ref{scatter}); because galaxy color distributions tend to be
asymmetric, we use the median absolute deviation (MAD) in our analysis
rather than assume a Gaussian distribution.  The MG1 sample has a
slope of $m= -0.017\pm 0.001$ (Table \ref{tbl.CMrelation slopes Ha<0}), and the slope stays essentially constant (approximately $m=-0.020 \pm0.003$)
in all 20 redshift bins.  The normalization of the CM relation
increases with increasing redshift due to the stretching and shifting
of the galaxy SED in the observed filters.  The errors associated with
fitting the CM relation also increase with redshift due to the smaller
luminosity range at higher redshift (see Figure \ref{fig.Absolute
magnitudes cut-off}).  This is true for all three of our Main Galaxy
samples.

MG1 has an observed color scatter of only
$\sigma_{\tiny{\textnormal{OBS}}}=0.0292 \pm0.0006$ (MAD value); here
we measure the error for $\sigma_{\tiny{\textnormal{OBS}}}$ using the
bootstrap resampling method ($e.g.$ \citealt{Efron.1979.Bootstrap};
see \S\ref{scatter}).  The color scatter is real, i.e. not dominated by
noise: the intrinsic scatter is essentially identical at
$\sigma_{INT}=0.0283\pm0.0006$ (Table \ref{tbl.CMrelation slopes Ha<0};
see \S\ref{scatter}).  Because the MG1 sample shows such a well-defined red
sequence in each redshift bin, we will later use its small color
scatter to exclude blue galaxies
[$\Delta(g'-r')>-3\sigma_{\tiny{\textnormal{OBS}}}$] in MG3.

\begin{figure}
\epsscale{1.0}
\plotone{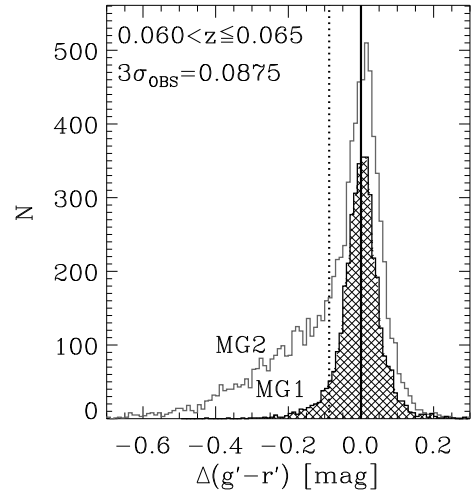}
\caption{Color deviation $\vartriangle$($g'-r'$) of MG1 (hatched
distribution) and MG2 (open) in the redshift range 0.060 $<$ $z$
$\leq$ 0.065 (see Figure \ref{fig.CMrelation slopes Ha+Sersic}); we
use the CM relation defined by MG1 to measure the color deviation for
both samples.  The dotted line 
is the 3$\sigma_{\tiny{\textnormal{OBS}}}$ (MAD) limit from MG1, and
we use it to exclude bluer galaxies from the MG2 (the S\'{e}rsic
sample) sample to define the MG3 sample. Note that for a Gaussian
distribution, 
$2\sigma\sim3\sigma_{\tiny{\textnormal{MAD}}}$.}
\label{fig.Histogram Ha+Sersic}
\end{figure}

The MG2 sample is shown in the middle panel of Figure
\ref{fig.CMrelation slopes Ha+Sersic} and its CM values are listed in
Table \ref{tbl.CMrelation slopes Sersic}. The CM relation for MG2 is
identical to that of MG1.  However, the MG2 sample has a measurably
larger observed color scatter of $\sigma_{\tiny{\textnormal{OBS}}}$ =
0.0646 (MAD); this is more than twice the color scatter of MG1. MG2's
wider range in color is due to a large number of blue galaxies: about
35\% of the MG2 galaxies lie below the color-cut defined by the MG1
sample (dotted line in Figure \ref{fig.CMrelation slopes Ha+Sersic}).

The difference in the color distribution between MG1 and MG2 are
especially striking in Figure \ref{fig.Histogram Ha+Sersic} which
shows the color deviation $\Delta(g'-r')$ for both samples.  MG1 has a
narrow and symmetric distribution, and less than 7\% of its galaxies
have $\Delta(g'-r')< -3\sigma_{\tiny{\textnormal{OBS}}}$.  In
comparison, 35\% of the galaxies in MG2 form an extended tail of blue
galaxies.

The color distribution for our third sample, MG3, is shown in the
lower panel of Figure \ref{fig.CMrelation slopes Ha+Sersic} and its CM
values are listed in Table \ref{tbl.CMrelation slopes
Sersic+color-cut}.  The CM relation in MG3 is identical to MG1 and
MG2, and the observed color scatter of
$\sigma_{\tiny{\textnormal{OBS}}}\sim0.0319$ is only slightly larger
than in MG1.

To summarize, the color scatter in MG2 (S\'{e}rsic $n >$ 2) is
significantly higher than in MG1 and MG3; this is true in each
redshift bin.  From the color distributions, it is clear that MG1
selects the galaxy sample with the most uniformly red and narrowest
range in color.  The CM relations in the three samples are consistent
with each other and with the slope of $m= -0.022$ measured by
\cite{Hogg.et.al.2004.redSequenceWithOverdensity} using SDSS galaxies
with S\'{e}rsic $n >$ 2.  Because we compare the three samples in each
redshift bin, our results are not affected by any bias introduced by
the $3''$ fiber.  The fact that the CM relation does not evolve across
our redshift range also indicates that aperture bias does not affect
our overall results when considering the entire redshift range
($0.05<z<0.15$).

\subsubsection{MG: Fraction of Passive Galaxies}
\label{subsec.Completeness}

From Tables \ref{tbl.CMrelation slopes Ha<0} - \ref{tbl.CMrelation
slopes Sersic+color-cut}, we see that the fraction of galaxies in each
sample varies significantly.  The MG1 sample is the most restrictive:
it contains only 22-37\% of the galaxies in each redshift bin (see
Table \ref{tbl.CMrelation slopes Ha<0}).  In comparison, both the MG2
and MG3 fractions are measurably higher at 51-79\% and 32-52\%,
respectively (Tables \ref{tbl.CMrelation slopes Sersic} \&
\ref{tbl.CMrelation slopes Sersic+color-cut}).  In all three samples,
the fraction increases with increasing redshift due to the higher
luminosity limit, $i.e.$ only the most luminous galaxies are in the
sample at higher redshift and these are predominantly passive systems
(\citealt{Feulner2005.SFRandGalsMass, Feulner2006.SFRandGalsMass}).

\begin{figure}
\epsscale{1.0}
\plotone{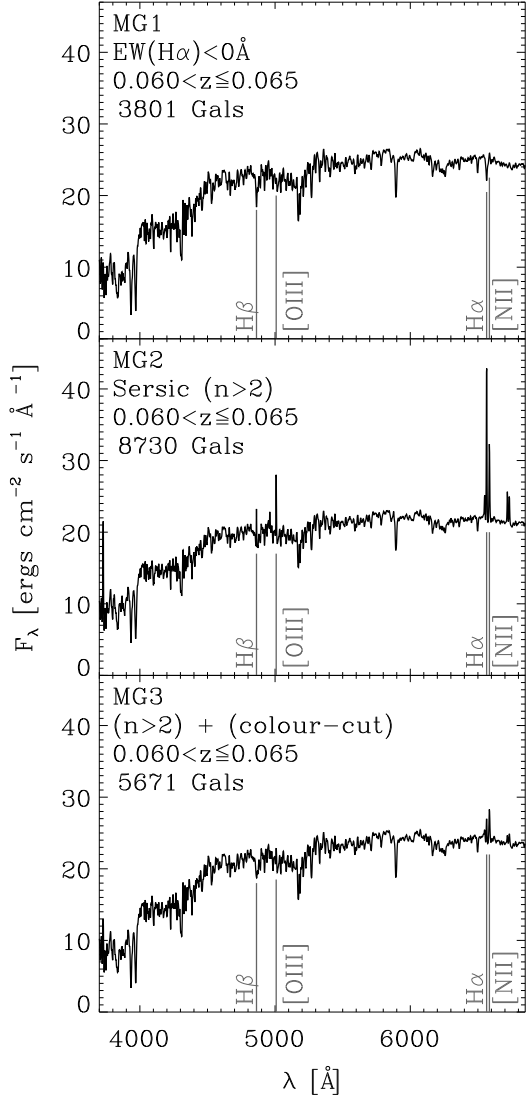}
\caption{Luminosity-weighted stacked spectra (rest-frame) for MG1 (top
panel), MG2 (middle panel), and MG3 (bottom panel) for galaxies at
$0.060<z\leq<0.065$; the spectral 
lines used for the BPT diagnostic are included. The
stacked spectrum for MG1 shows no emission in any of the noted lines
and indicates that MG1 contains the oldest galaxy population. In
contrast, both MG2 and MG3 show measurable line emission in
H$\alpha$.}
\label{fig.average Spectrum Ha+Sersic}
\end{figure}

We find that almost all of the galaxies in MG1 ($>92$\%) are also in
MG2 and MG3, but the reverse is not true: only 43\% of the galaxies in
MG2 have H$\alpha$ absorption, and 62\% in MG3.  Although MG2 and MG3
have more galaxies than MG1, they both include emission-line (active)
systems.  Luminosity-weighted stacked spectra for each sample (Figure
\ref{fig.average Spectrum Ha+Sersic}) confirm that MG1 does not show
any measurable emission and has a high D4000 value of 1.73 while both
MG2 and MG3 include active galaxies with lower D4000 values of 1.54 and
1.71, respectively.  These results combined with the color analysis
suggests that the H$\alpha$ absorption-line criterion may be selecting
only the ``core'' galaxies of the red sequence.

\subsubsection{MG: Star Formation vs. AGN}
\label{subsec.AGN fraction}

\begin{figure}
\epsscale{1.0}
\plotone{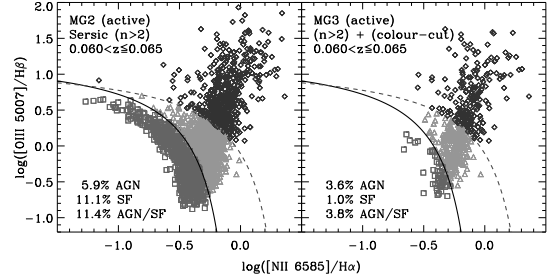}
\caption{The BPT-diagrams of the active galaxies in MG2 (left panel)
and MG3 (right panel) in the redshift range of 0.060 $<$ $z$ $\leq$
0.065. The black solid curve represents the separation curve of
\cite{Kauffmann.et.al.2003.BPTplot} and the grey dotted curve the
separation curve of \cite{Kewley.et.al.2001.BPTplot}.  The squares
represent purely star-forming (SF) galaxies, the triangles hybrid
AGN/SF galaxies, and the diamonds AGN dominated galaxies; the
galaxy fractions for these three categories are in the lower left
corner of each panel and are calculated with respect to the total
galaxy number in each sample.  Applying the color-cut (MG3) reduces
the fraction of active galaxies significantly, but there is still
measurable contamination.}
\label{fig.BPTplot Sersic+Sersic-cut}
\end{figure}

Having found that both the MG2 and MG3 samples contain active
galaxies, we examine whether the activity is due to ongoing star
formation or AGN.  The spectral range of our stacked spectra include
H$\alpha$, H$\beta$, [NII], and [OIII]; these four lines are often
used to separate star-forming galaxies from those dominated by
(optically-identified) AGN (\citealt{Baldwin.et.al.1981.BPTplot}).  Note
that we only consider the MG2 and MG3 samples here because the MG1
sample, by definition, does not have any emission lines (see
Fig. \ref{fig.average Spectrum Ha+Sersic}).

Figure \ref{fig.BPTplot Sersic+Sersic-cut} shows the BPT diagrams for
the MG2 (left panel) and MG3 (right panel) samples in the redshift bin
($0.060<z\leq0.065$). The active galaxies are separated into three
categories using the curves from
\cite{Kauffmann.et.al.2003.BPTplot} and
\cite{Kewley.et.al.2001.BPTplot}: 1) star-forming; 2) AGN; and 3)
combination of SF/AGN.  The SF/AGN galaxies lie between the two curves
and are neither clearly star-forming nor purely AGN.  The fractions of
galaxies in each category are calculated with respect to the total
galaxy number $N_{gal}$ (see Tables \ref{tbl.CMrelation slopes Sersic} \&
\ref{tbl.CMrelation slopes Sersic+color-cut}) and are listed in
Tables \ref{tbl.BPT Sersic} \& \ref{tbl.BPT Sersic+color-cut}.

In our lowest redshift bin, 31\% and 9\% of the galaxies in the MG2
and MG3 samples, respectively, are active (Tables \ref{tbl.BPT Sersic}
\& \ref{tbl.BPT Sersic+color-cut}).  However, the fraction of active
galaxies decreases with increasing redshift because the luminosity
range in the Main Galaxy samples also decreases with increasing
redshift (see Figure \ref{fig.Absolute magnitudes cut-off}).  The
effect is strongest in MG2 where the fraction of star-forming galaxies
drops from 15\% in the lowest redshift bin which has a magnitude limit
of $M_r<-18.5$ mags to 2\% in the highest redshift bin with
$M_r<-21.0$ mags (Table \ref{tbl.BPT Sersic}).  In comparison, the
fraction of AGN in the MG2 sample only decreases from 6\% to 4\%, and
the fraction of SF/AGN galaxies decreases by nearly half (11\% to
6\%).  These trends indicate that the galaxies with AGN are on average
brighter than those with ongoing star-formation (see also \citealt{Sadler.et.al1999.AGNsAndSFgals, Sadler.et.al2002.AGNsBrighterThanSF}).

MG3 has a lower fraction of active galaxies than MG2 because the
color-cut removes many of the SF and AGN/SF galaxies
(Fig. \ref{fig.BPTplot Sersic+Sersic-cut} \& Table \ref{tbl.BPT
Sersic+color-cut}).  The AGN fraction in MG3 is lower than that in MG2
(3.6\% vs. 6\%; $0.060<z\leq0.065$).  Note that in MG3, the AGN
fraction (3.6\%) is measurably higher than its SF fraction (1\%; Table
\ref{tbl.BPT Sersic+color-cut}).  These results show that early-type
($n>2$) galaxies with AGN tend to be on the red sequence, and that
most early-type galaxies with ongoing star formation are blue.  As an
interesting sidenote, we also find that the AGN in MG3 are not as red
as the MG3 galaxies with EW(H$\alpha$)$<0$\AA; like
\cite{Schawinski.et.al.2009GalaxyZooRedSequence}, we find early-type
galaxies with AGN lie closer to the ``green valley''.  Further study
of these early-types with AGN may place useful constraints on the link
between AGN and the quenching of star formation, but such an analysis
is beyond the scope of this paper.

\subsubsection{MG: D4000 Index \& Stellar Age}
\label{subsec.Median Ages}

\begin{figure}
\epsscale{1.0}
\plotone{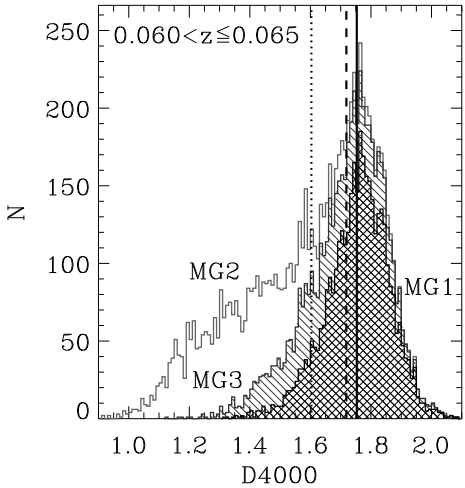}
\caption{The D4000 distribution of MG1 (hatched), MG2 (open), and MG3
(diagonal shaded). The average D4000 value of each sample is indicated
by the solid  (MG1), dashed (MG2), and dotted (MG3)
lines. MG1 has the highest average D4000 value and a symmetric
distribution. In contrast, MG2 and MG3 have extended tails towards
lower D4000 values, i.e. they both include younger galaxies.}
\label{fig.D4000 plot of Ha<0}
\end{figure}

The D4000 index increases as a stellar population ages
(\citealt{Bruzual.1983.D4000value, Kauffmann.et.al.2003.Dn4000value})
and can be used to compare the relative ages of the three Main Galaxy
samples.  Table \ref{tbl.averageD4000 values} lists the median D4000
value in each redshift bin for MG1, MG2, and MG3, and Figure
\ref{fig.D4000 plot of Ha<0} shows how their D4000 distributions
differ at ($0.060<z\leq0.065$): MG1 has a narrow, symmetric
distribution centered at D4000=1.75, but both MG2 and MG3 have
asymmetric distributions with long tails extending to lower D4000
values.

In detail, more than 85\% of the galaxies in MG1 have D4000$>1.60$,
and none have D4000$<1.30$.  In contrast, about 10\% of the MG2
galaxies have D4000$<1.30$, and MG2 has a lower median D4000 value of
only 1.60. In MG3, the color-cut does remove nearly all galaxies with
D4000$<1.30$, but MG3's median D4000 of 1.72 is still lower than in
MG1.  As a check, we also measure D4000 in the stacked spectrum for
each galaxy sample (see Figure \ref{fig.average Spectrum Ha+Sersic})
and find that the values are virtually identical to our results using
the D4000 distributions.

To estimate the stellar age corresponding to the median D4000 values in
our three samples, we use the stellar synthesis model from
\cite{Bruzual.Charlot.2003.BC03}.  For simplicity, we assume:

\begin{itemize}
\item solar metallicity (Z = 0.02)
\item Salpeter initial mass function (IMF)
\item single instantaneous starburst
\label{list.BC03}
\end{itemize}

\noindent Because we are only interested in knowing the relative age
differences between the samples, we use a single metallicity and
simple starburst model.  The stellar ages corresponding to the median
D4000 value in MG1, MG2, and MG3 are listed in Table
\ref{tbl.averageD4000 values}.  As a check, we also estimate the ages
from the stacked spectrum for each sample and find that the ages
measured by the two methods are consistent.

The high median D4000 value and small scatter of the H$\alpha$
selected galaxies (MG1) means that these galaxies have highly uniform
and relatively old ages.  The difference between MG1 and MG2 (S\'ersic
$n>2$) is striking: the median age in MG1 is about double that in MG2.
In comparison, the red early-type sample (MG3) has a median age closer
to, but never as old as, the MG1 sample (Tables \ref{tbl.averageD4000
values}).  Note that the median age in all three samples increases
with increasing redshift due to the brighter magnitude limit, $i.e.$
more luminous galaxies have older stellar populations and higher D4000
values (\citealt{Gallazzi.et.al2006.ageMetalGalsLocalUniverseSDSS}).

\subsubsection{MG: S\'{e}rsic index vs. D4000 }
\label{subsec.Sersic and D4000}

\begin{figure}
\epsscale{1.0}
\plotone{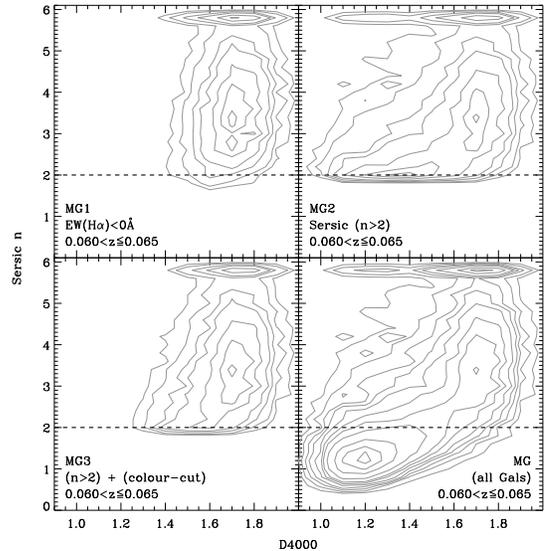}
\caption{The distribution of S\'{e}rsic index $n$ versus D4000 for MG1
(upper left), MG2 (upper right), MG3 (bottom left), and all galaxies
(bottom right); all distributions are for the same redshift
range. The dashed line inside each panel represents the S\'{e}rsic
index cut-off of $n$ = 2 used to select MG2 and MG3. The overdensity
at S\'{e}rsic n $\sim$ 6 is based on the NYU catalog's imposed
threshold where every galaxy with n $>$ 6 has a listed S\'{e}rsic
index of $n$ = 6. MG1 (the H$\alpha$ sample) consists nearly
exclusively of bulge dominated galaxies with a S\'{e}rsic index $n >$
2.}
\label{fig.D4000.vs.Sersic Sample I}
\end{figure}

Morphology, as quantified by the S\'ersic index, is often used to
identify passive galaxies because they also tend to be early-type
systems ($n>2$, $e.g.$
\citealt{Hogg.et.al.2004.redSequenceWithOverdensity}). We test this
assumption by comparing D4000, a robust tracer of stellar age, to the
S\'ersic indices from the NYU VAGC.  In Figure
\ref{fig.D4000.vs.Sersic Sample I} (bottom right), we plot D4000
versus S\'ersic index for all the galaxies at 0.06 $< z \leq$ 0.065 in
the Main Galaxy sample; the overdensity at $n=6$ is due to the NYU
VAGC assigning a value of $n=6$ for any galaxy with a measured
S\'ersic index greater than 6.

The distribution for all galaxies is bimodal with a concentration at
low D4000 ($<1.4$) and low $n$ ($<2$), and another at high D4000
($>1.6$) and high $n$ ($>2$).  Note that the color-magnitude diagram
has a similar bimodal distribution
(\citealt{Baldry.et.al2004.BimodalColor-MagnitudeDistribution}).
However, the correlation between S\'ersic index and D4000 is quite
broad, $e.g.$ a galaxy with $n=2$ can have a D4000 value from 1.0 to
1.8.

In comparison, the S\'ersic and D4000 distribution of the MG1
(H$\alpha$) sample is strikingly narrow (Figure
\ref{fig.D4000.vs.Sersic Sample I}, top left).  Even though we did
not select on morphology, virtually all of the MG1 galaxies have
$n>2$.  As noted earlier, their D4000 distribution includes only
values $>1.4$.

In MG2 and MG3 where we do select using morphology ($n>2$), the D4000
distribution is visibly broader than in MG1 (Figure
\ref{fig.D4000.vs.Sersic Sample I}).  The contamination by younger
galaxies (D4000$<1.4$) is particularly severe in the MG2 sample (top
right).  Applying the color-cut in the MG3 sample does remove the
galaxies with D4000$<1.3$ (bottom left panel), but MG3 still contains
a number of galaxies with $1.3<$D4000$<1.4$.


\subsection{Luminous Galaxy Sample}
\label{subsec.AbsMag Sample I,II,III}

Thus far we have only used the Main Galaxy sample which is divided
into 20 redshift bins where the magnitude limit increases with
redshift to ensure a complete sample in each redshift bin.  To test
the robustness of our results, we now apply an absolute magnitude
limit of $M_{r'}\leq-21.0$ set by our highest redshift bin to the
entire galaxy sample ($0.05<z\leq0.15$); this leaves us with
$\sim82,000$ galaxies (see Table \ref{tbl.Main galaxy sample}).  We
divide the Luminous Galaxy sample into the same three categories as
defined in Table \ref{tbl.Luminous galaxy sample}: LG1 (H$\alpha$
EW$<0$\AA); LG2 (S\'ersic $n>2$); and LG3 (S\'ersic $n>2$ and
color-cut).  We repeat our analysis with the Luminous Galaxy sample
and compare to our results from the Main Galaxy sample.  Note that
because of the larger redshift range, we now use absolute rest-frame
magnitudes and colors from the NYU VAGC.

\subsubsection{LG: Color-Magnitude Relation}

\begin{figure}
\epsscale{1.0}
\plotone{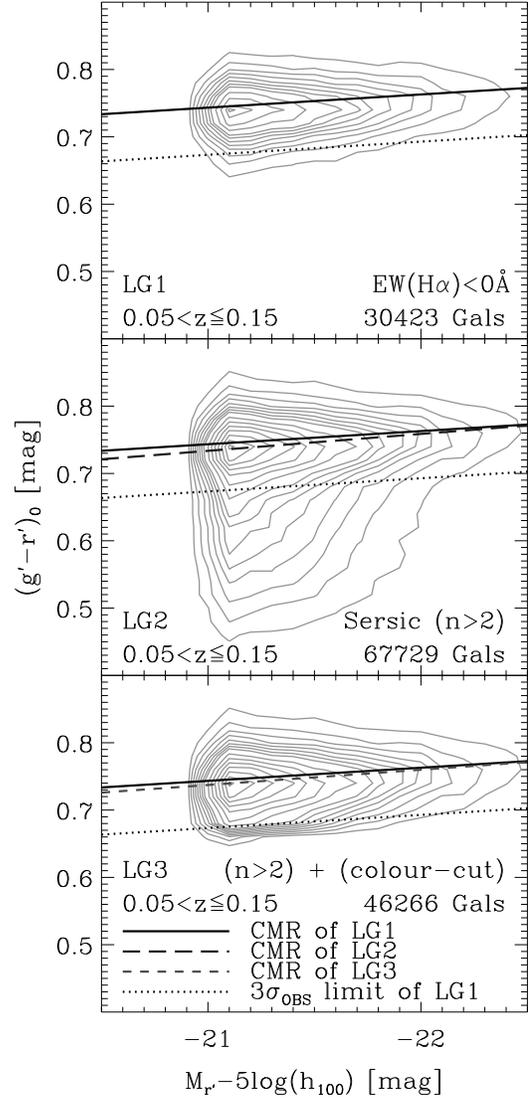}
\caption{Rest-frame color-magnitude diagrams for LG1 (H$\alpha$
passive; top panel), LG2 (S\'{e}rsic $n >$ 2; middle panel), and LG3 (S\'{e}rsic $n >$ 2 and
color-cut; bottom panel). We consider all galaxies at $0.05<z\leq0.15$ with $M_{r'}$
$<$ -21.0 (see Fig. \ref{fig.Absolute magnitudes cut-off}). The solid
line is the sigma-clipped least-squares fit to LG1 (shown in all panels), the long-dashed
line the fit to LG2, and the short-dashed line to LG3. The dotted line
shown in all three panels 
represents the 3$\sigma_{\tiny{\textnormal{OBS}}}$ (MAD) limit from LG1
(3$\sigma_{\tiny{\textnormal{OBS}}}$ = 0.0700).  Because of 
K-corrections, plotting the three galaxy samples in the rest-frame
results in a tighter red sequence than plotting
the samples in the observed frame (see Figure
\ref{fig.CMrelation slopes Ha+Sersic}).  However, the
H$\alpha$-criterion (LG1) still isolates the most uniformly red (old)
galaxy population.} 
\label{fig.AbsMag Ha<0}
\end{figure}

The color-magnitude diagrams for LG1, LG2, and LG3 are shown in
Figure \ref{fig.AbsMag Ha<0}, and their fitted color-magnitude
relations are listed in Table \ref{tbl.AbsMag CMrelations}.  As in our
earlier analysis, we use $3\sigma_{OBS}$ measured in LG1 for the
color-cut applied in LG3.  The CM relations for all three samples are
virtually identical to each other as well as to the Main Galaxy
sample; note that the normalizations are different between the Main
and Luminous Galaxy samples due to the change from apparent to
absolute magnitudes.

\begin{figure}
\epsscale{1.0}
\plotone{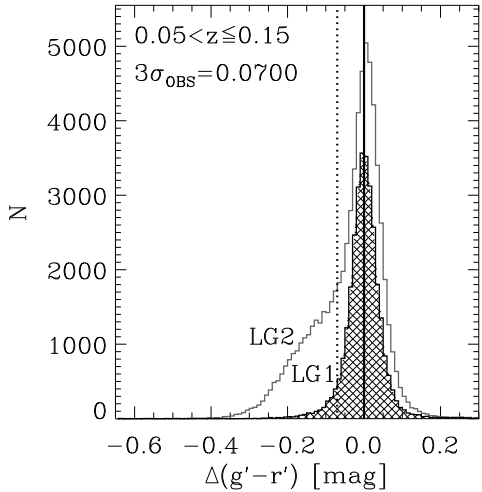}
\caption{Histograms of the rest-frame color residuals
$\vartriangle$($g'-r'$) of LG1 (hatched) and LG2 (open); we use the CM
relation from LG1 for both samples. The dashed line is the
3$\sigma_{\tiny{\textnormal{OBS}}}$ (MAD) of the LG1 sample used with
$n>2$ to define the LG3 sample. LG1 has a very peaked and symmetric
color distribution while LG2 has an extended tail towards bluer
(younger) galaxies. }
\label{fig.AbsMag Ha+Sersic} 
\end{figure}

The observed color scatter is lowest in LG1
($\sigma_{\tiny{\textnormal{OBS}}}$=0.023) while the scatter is twice
as large in LG2 ($\sigma_{\tiny{\textnormal{OBS}}}$=0.046).  LG1 has a
narrow and symmetric color distribution, but LG2's color deviation is
again clearly asymmetric with a tail of bluer galaxies (see Figure
\ref{fig.AbsMag Ha+Sersic}).  The color-cut in LG3 does reduce the
color scatter considerably to
$\sigma_{\tiny{\textnormal{OBS}}}$=0.025, but LG3 still contains a
significant number of red active galaxies (see Figure \ref{fig.AbsMag
BPTplots}, right panel).  Note that the color scatter in the Luminous
Galaxy samples are smaller than those in the Main Galaxy samples due
to the higher luminosity limit and the use of rest-frame colors.

The same is true when comparing the intrinsic color scatter of LG1,
LG2, and LG3 (Table \ref{tbl.Luminous galaxy sample}): the LG1 sample
has the lowest intrinsic color scatter.  To determine the intrinsic
color scatter, we correct the observed color scatter using the
photometric errors listed in the NYU VAGC (see \S\ref{scatter}).  In all
cases, the color scatter is primarily due to intrinsic variations in
the galaxy population and not to systematic errors.

\begin{figure}
\epsscale{1.0}
\plotone{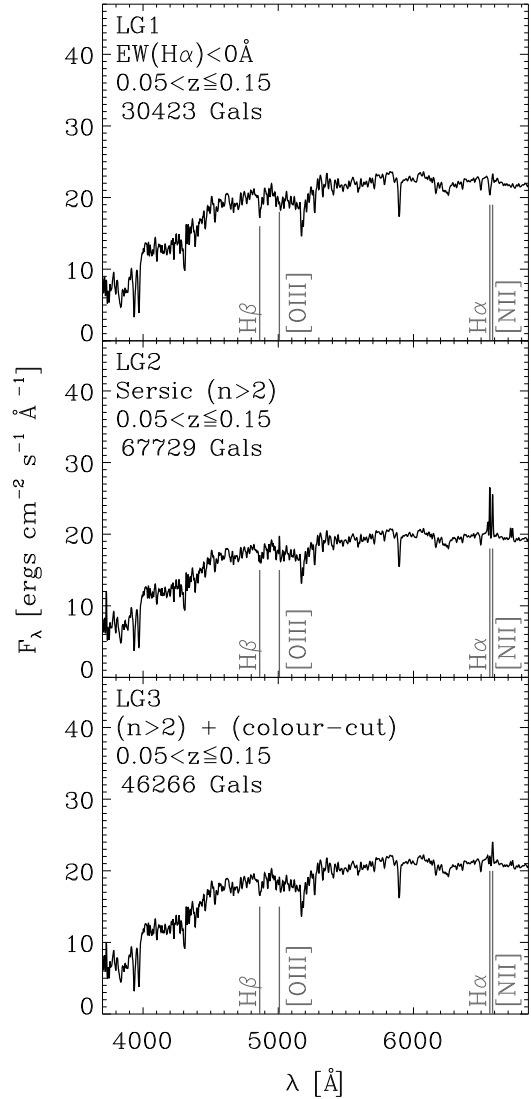}
\caption{Luminosity-weighted stacked spectra of LG1 (top panel), LG2
(mid panel), and LG3 (bottom panel). While the H$\alpha$ emission in LG2 and
LG3 are weaker than in the Main Galaxy samples,  only LG1 shows no
emission in any of the noted lines.  Thus even in the Luminous Galaxy
samples, selecting by H$\alpha$ is still the most effective criteria
for identifying a homogeneously old population.}
\label{fig.AbsMag SpecMultiplot}
\end{figure}

As in the Main Galaxy sample, LG1 (H$\alpha$-selected) has the fewest
galaxies while LG2 (morphology-selected) has the most (37\% vs. 83\%),
and LG3 is in between (57\%).  However, the fraction of passive
galaxies in LG2 and LG3 are only 44\% and 62\% respectively, $i.e.$
both LG2 and LG3 have a significant number of emission-line galaxies.
The average luminosity-weighted stacked spectra for LG1, LG2 and LG3
confirm that L1 is composed of completely passive galaxies (Figure
\ref{fig.AbsMag SpecMultiplot}) while both LG2 and LG3 include active
systems; the nearly equal emission strength in H$\alpha$ and [NII]
indicate that the active galaxies in LG2 and LG3 are mostly AGN.  As
noted earlier, the predominance of AGN rather than ongoing star
formation is due to the brighter luminosity limit ($M_{r'}<-21.0$) in
the Luminous Galaxy sample: the fainter galaxies that are excluded
tend to be star-forming systems while the more luminous galaxies tend
to have AGN ($e.g.$ see Table \ref{tbl.BPT Sersic}).

\subsubsection{LG: Star Formation vs. AGN}

\begin{figure}
\epsscale{1.0}
\plotone{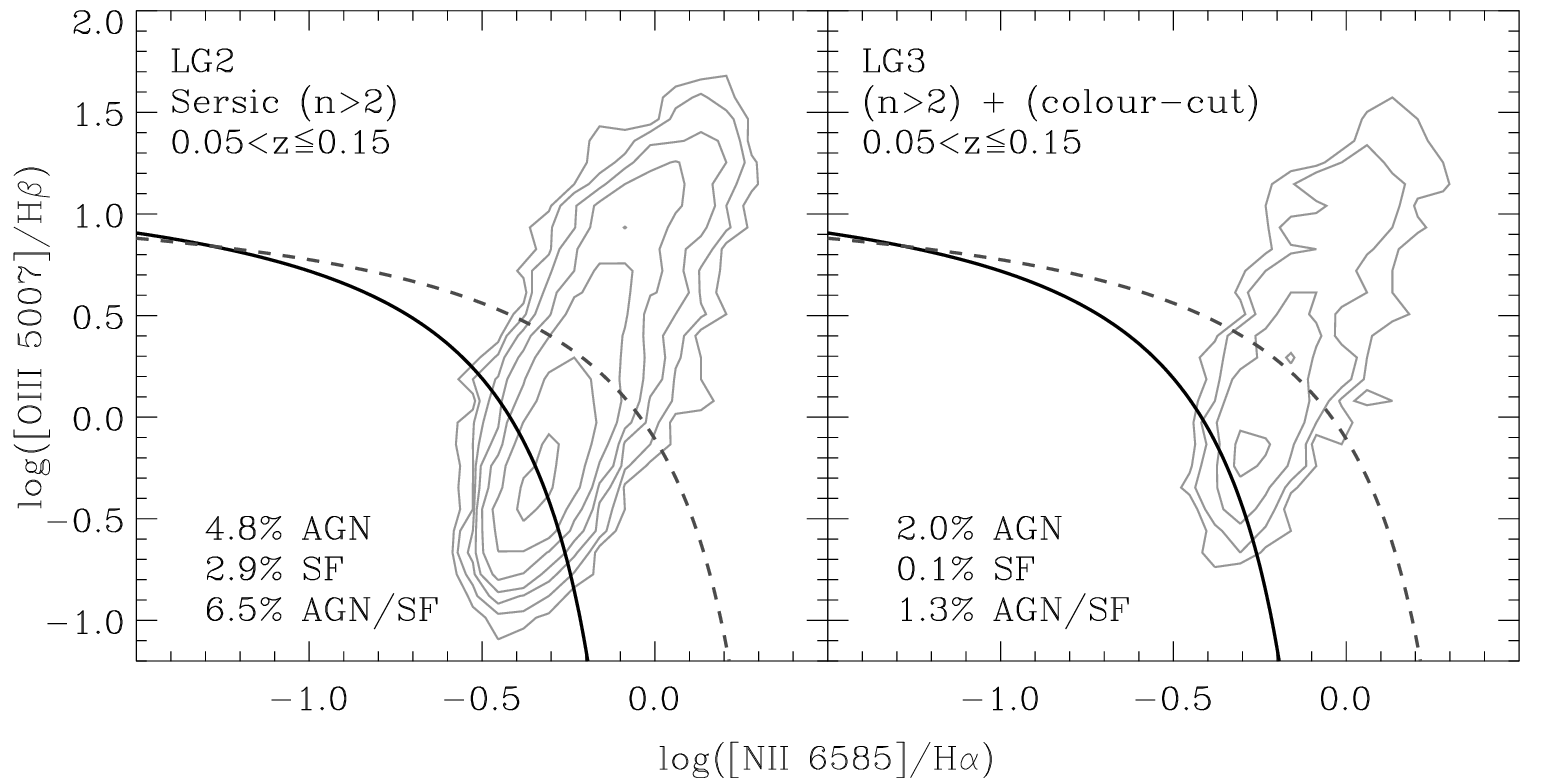}
\caption{The BPT-diagrams of the active galaxies in LG2 (left panel;
14.2\% of all LG2 galaxies) and LG3 (right panel; 3.4\% of all LG3
galaxies). The solid curve represents the separation curve of
\cite{Kauffmann.et.al.2003.BPTplot} and the dotted curve the
separation curve of \cite{Kewley.et.al.2001.BPTplot}. The relative
fraction of active galaxies in the Luminous Galaxy samples is lower
than in the Main Galaxy samples (Figure \ref{fig.BPTplot
Sersic+Sersic-cut}) due to the higher luminosity limit of the LG
samples. However, both LG2 and LG3 still have measurable contributions
from SF/AGN galaxies. }
\label{fig.AbsMag BPTplots}
\end{figure}

We use the BPT diagram again to separate purely star-forming galaxies
from those with AGN; this is only possible with the LG2 and LG3
samples because they both have galaxies with emission lines while the
LG1 sample does not (Figure \ref{fig.AbsMag BPTplots}).  Of the
galaxies in LG2, 14\% are active systems.  The active fraction is
lower in LG3 at only 3\%.  In both LG2 and LG3, the majority of the
active galaxies have AGN.  This reinforces our earlier results showing
that the bright active galaxies tend to have AGN rather than ongoing
star formation ($e.g.$ see Tables \ref{tbl.BPT Sersic} - \ref{tbl.BPT
Sersic+color-cut}).

\subsubsection{LG: D4000 Index \& Stellar Age}
\label{subsec.Age of AbsMag Samples}

\begin{figure}
\epsscale{1.0}
\plotone{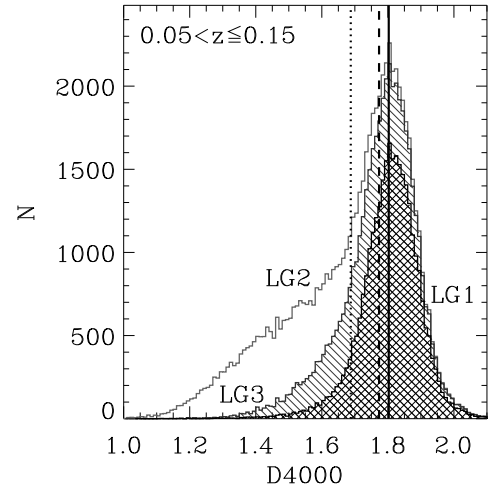}
\caption{The D4000 distribution of LG1 (hatched), LG2 (open), and LG3
(diagonal shaded). The average D4000 value for each sample is
indicated by the solid (LG1), dashed (LG2), and dotted (LG3) lines. As
in the Main Galaxy samples, LG1 has the highest average D4000 value
and a symmetric distribution while both LG2 and LG3 have extended
tails towards lower D4000 values, i.e. they both include younger
galaxies.}
\label{fig.AbsMag Ha+Sersic+Cut}
\end{figure}

The D4000 distributions of LG1, LG2, and LG3 are shown in Figure
\ref{fig.AbsMag Ha+Sersic+Cut}: the LG1 sample has the narrowest and
most symmetric D4000 distribution, and it has the highest mean value
of 1.80.  The LG2 sample has the widest D4000 distribution with an
asymmetric tail towards lower values; its mean value is only 1.69.
Because the color-cut in LG3 tends to remove star-forming galaxies
that have lower D4000 indices, its mean D4000 value is higher at 1.77.
As a check, we also measure the D4000 indices in the stacked spectra
(Figure \ref{fig.AbsMag SpecMultiplot}) and find the results to be
consistent.

Using the same BC03 starburst model as in \S\ref{subsec.Median
Ages}, we determine the stellar ages as measured by the D4000 index.
Because LG1 has the highest mean D4000 value, it has the oldest mean
stellar age of 4.5 Gyr while LG2 has a mean age of only 3.0 Gyr.  Thus
even in a luminous galaxy sample, the H$\alpha$ selection continues to
be the most effective criteria for isolating a uniformly quiescent
galaxy population.

The Luminous Galaxy samples have higher mean D4000 indices and thus
older mean stellar ages relative to the Main Galaxy samples.  The Main
Galaxy samples are younger due to the inclusion of fainter galaxies
that tend to have more ongoing star formation.  In other words, more
luminous galaxies tend to have older stellar populations, as found in
previous studies
(\citealt{Gallazzi.et.al2006.ageMetalGalsLocalUniverseSDSS,
Graves.et.al.2007.RedSequence}).

\subsubsection{LG: S\'ersic index vs. D4000}

\begin{figure}
\epsscale{1.0}
\plotone{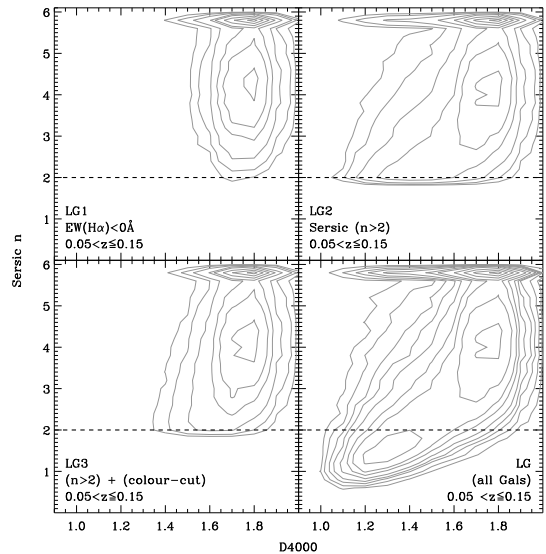}
\caption{The distribution of S\'{e}rsic index $n$ versus D4000 for 
LG1 (upper left panel), LG2 (upper right), LG3
(lower left), and all galaxies (bottom right).  As in the Main Galaxy
samples, the H$\alpha$-selected Luminous Galaxy sample (LG1) consists
nearly exclusively of bulge-dominated galaxies ($n\geq2$) and high
D4000 values.} 
\label{fig.AbsMag D4000.vs.Sersic}
\end{figure}

Finally, we compare how well S\'ersic index correlates with D4000 in
the Luminous Galaxy samples.  The results for the Luminous Galaxy
samples (Figure \ref{fig.AbsMag D4000.vs.Sersic}) are essentially the
same as in the Main Galaxy samples (Figure \ref{fig.D4000.vs.Sersic
Sample I}): When considering the entire Luminous Galaxy sample, the
distribution is bimodal.  However, the H$\alpha$ selection (LG1)
contains only galaxies with $n>2$ and D4000$>1.5$.  Both of the
morphologically-selected samples (LG2 \& LG3) have wider
distributions.

In summary, we confirm that using only H$\alpha$ is as effective in
the Luminous Galaxy sample as in the Main Galaxy sample at isolating a
galaxy population that is highly uniform in color (red), D4000 index
(high), and morphology ($n>2$).

\section{Discussion}
\label{sec.Discussion}

The properties of the oldest galaxies at any redshift provide strong
constraints on both galaxy formation models
(\citealt{Bower.et.al2006.GalaxyFormation,
Font.et.al2008.ColorSatelliteGals}) and cosmological parameters,
$e.g.$ how the Hubble parameter evolves with redshift
(\citealt{Stern.et.al2009.H(z)Measurements}).  Specifically, the
observed color distribution of a uniformly aged population is a
straight-forward test of galaxy formation models, and stellar
population models must be able to reproduce the observed age
distribution and remain consistent with the universe's age at any
redshift.  However, how to identify a galaxy population with such
homogeneous mean stellar ages continues to be debated because of
mixed success using morphologically and/or color-selected samples:
these selection techniques are unable to exclude a number of galaxies
with ongoing star formation, particularly at $L<L^{\ast}$
(\citealt{Yan.et.al.2006.OIIinRedSequence}).

Given the importance of identifying a uniformly old galaxy population,
especially at higher redshift, we have introduced a new method that
relies solely on the H$\alpha$ line.  We find that galaxies with
EW(H$\alpha)<0$\AA\ have a narrow and symmetric color distribution
(Figs. \ref{fig.CMrelation slopes Ha+Sersic} \& \ref{fig.Histogram
Ha+Sersic}) with an observed color scatter of
$\sigma_{\tiny{\textnormal{OBS}}}(g'-r')=0.029$ (MAD) in our lowest
redshift bin ($z\sim0.05$) and 0.038 in our highest redshift bin
($z\sim0.15$).  When we remove systematic errors to obtain the
intrinsic color scatter, we find that the observed color scatter is
dominated by real variations in the galaxy population with
$\sigma_{\tiny{\textnormal{INT}}}=0.0287$ ($z\sim0.05$) to
$\sigma_{\tiny{\textnormal{INT}}}=0.0357$ ($z\sim0.15$); these values
are for photometry in the observed frame, i.e. without K-corrections.

The slope in the color-magnitude relation of $m=-0.02$ is constant
throughout our 20 redshift bins (Table \ref{tbl.CMrelation slopes
Ha<0}).  Our results do not change even when considering only the most
luminous ($M_{r'}<-21.0$) galaxies across the entire redshift range
(Figs. \ref{fig.AbsMag Ha<0} \& \ref{fig.AbsMag Ha+Sersic}): the slope
in the CM relation is the same, and the absolute color scatter is
$\sigma_{\tiny{\textnormal{OBS}}}(g'-r')=0.023$.  We also find that
all of the H$\alpha$ selected galaxies have morphological S\'ersic
indices of $n\geqslant2$, $i.e.$ they are all bulge-dominated systems
\citep{Sersic.1968.sersicIndex,Driver2006.SersicMorphology,Wel2008.SersicMorphology}.

As a check of the H$\alpha$ selected samples, we determine the average
luminosity-weighted spectrum in each redshift bin for the Main and
Luminous Galaxy samples (Figs. \ref{fig.average Spectrum Ha+Sersic} \&
\ref{fig.AbsMag SpecMultiplot}).  The stacked spectra have no emission
lines and D4000, a measure of the continuum break at 4000\AA,
increases from 1.72 at $z\sim0.05$ to 1.80 at $z\sim0.15$ (Table
\ref{tbl.averageD4000 values}).  Using a BC03 single starburst model
(solar metallicity), the corresponding mean stellar ages of the
H$\alpha$-selected sample is 3 to 4.25 Gyr.  Note that the D4000
distributions of the H$\alpha$ selected samples are narrow and
symmetric, and all of the galaxies have D4000$>1.3$
(Figs. \ref{fig.D4000 plot of Ha<0} \& \ref{fig.AbsMag
Ha+Sersic+Cut}).

We quantify the effectiveness of our H$\alpha$-selection by comparing
to both morphologically and color-selected samples.  Following
\cite{Hogg.et.al.2004.redSequenceWithOverdensity}, we select galaxies
with early-type morphologies by using S\'ersic $n>2$.  We find that
while the slope of the CM relation is the same as in the
H$\alpha$-selected samples, the distribution in both color and D4000
index is significantly wider with asymmetric tails towards bluer
galaxies with lower D4000 indices (Figs. \ref{fig.CMrelation slopes
Ha+Sersic}, \ref{fig.Histogram Ha+Sersic}, \ref{fig.D4000 plot of
Ha<0}, \& \ref{fig.D4000.vs.Sersic Sample I}).  About 35\% of the
($n>2$) galaxies have $\Delta(g'-r')<-3\sigma_{\tiny{\textnormal{OBS}}}$ (where
$\sigma_{\tiny{\textnormal{OBS}}}$ defined by the H$\alpha$ samples), and their observed
color scatter is twice that of the H$\alpha$-selected samples (Tables
\ref{tbl.CMrelation slopes Sersic} \& \ref{tbl.CMrelation slopes
Sersic+color-cut}).

The morphologically-selected samples have a significant fraction of
active galaxies, and their stacked spectra have strong emission lines
(Figs. \ref{fig.average Spectrum Ha+Sersic} \& \ref{fig.BPTplot
Sersic+Sersic-cut}).  Due to the increasing magnitude limit in the
redshift bins, the active fraction decreases from 32\% at $z\sim0.05$
to 13\% at $z\sim0.15$ (Table \ref{tbl.BPT Sersic} \& \ref{tbl.BPT
Sersic+color-cut}).  The relative fraction of galaxies with ongoing
star formation versus those with AGN also depends on the luminosity
limit: active galaxies that are fainter than $M_{r'}=-20.0$ tend to be
dominated by star formation instead of AGN.  Because of these
star-forming systems, the ($n>2$) samples have lower mean D4000
indices of 1.51--1.69 ($0.05<z\leq0.15$; Table \ref{tbl.averageD4000
values}); the corresponding stellar ages (1.20 to 3.00 Gyr) are about
half that of the H$\alpha$-selected samples (Table
\ref{tbl.averageD4000 values}).

Having established that the H$\alpha$-selected sample is more
homogeneous in age than the morphologically-selected sample, we also test
if applying a color-cut to the $n>2$ sample eliminates the
emission-line galaxies.  Using the observed color scatter
$\sigma_{\tiny{\textnormal{OBS}}}$ determined from the H$\alpha$
samples, we exclude all $n>2$ galaxies that are bluer than
$3\sigma_{\tiny{\textnormal{OBS}}}$.  We find that while the color-cut
does reduce the active fraction considerably, the galaxy sample is
still not as uniform in color and D4000 (stellar age) as the H$\alpha$
samples (Figs. \ref{fig.CMrelation slopes Ha+Sersic},
\ref{fig.Histogram Ha+Sersic}, \& \ref{fig.D4000 plot of Ha<0}; Table
\ref{tbl.averageD4000 values}).

We find that our single selection criterion of EW(H$\alpha)<0$\AA\ is
more effective than using S\'ersic index and/or color cuts to identify
a galaxy population that is homogeneous in color (red), D4000 index
(high), and morphology ($n>2$).  This is true in both our Main Galaxy
sample where we use a varying magnitude limit that depends on
redshift, and in the Luminous Galaxy sample where we apply an absolute
magnitude limit of $M_{r'}<-21.0$.  Note that even in the LGS where
the galaxies are dominated by older stellar populations, there is
still measurable contamination from active systems when selecting by
S\'ersic index and/or color.

Our ability to measure H$\alpha$ emission is not as strongly dependent
on redshift as measuring S\'ersic index: while quantifying galaxy
morphology at $z>0.5$ requires imaging resolution of $<0.1''$,
identifying galaxies with EW(H$\alpha$)$<0$\AA\ can be done with even
low signal-to-noise ($<2$) spectra.  The main limitation of using
H$\alpha$ is that the spectral line moves into the infrared for
galaxies at $z>0.4$.  However, the recent development of multi-object
infrared spectrographs enables us to select with H$\alpha$ to
$z\sim2$.

\section{Conclusion}
\label{sec.Conclusion}

By combining the SDSS DR6 with the NYU VAGC, we obtain a sample of
over 180,000 galaxies at $0.05<z<0.15$ with $u'g'r'i'z'$ photometry,
S\'ersic index, and measured spectral indices (Main Galaxy sample; MG).
In an effort to find the most effective observational method of
identifying a truly quiescent red sequence, we select
galaxies solely by their H$\alpha$ line.  We compare the galaxies with
H$\alpha$ equivalent widths of $<0$\AA, $i.e.$ in absorption, to
galaxy samples selected using the more common methods of morphology
(S\'ersic index) and color.  Our work complements earlier studies
(\citealt{Gallazzi.et.al2006.ageMetalGalsLocalUniverseSDSS,Graves.et.al.2009.RedSequence})
by rigorously testing which selection criteria identifies the most
homogeneously aged red sequence.

We split the Main Galaxy sample into 20 redshift bins that have an
increasing luminosity limit with redshift; this ensures that we are
complete in each magnitude bin ($\Delta z=0.005$) and, by comparing
the samples within each redshift bin, we also minimize any bias due to
the $3''$ fiber aperture.  We repeat our analysis using a Luminous
Galaxy (LG) sample where we apply an absolute magnitude limit of
$M_{r'}<-21.0$ over the entire redshift range.

We measure the slope of the color-magnitude relation to be $m=-0.02$
which is consistent with the slope determined by
\cite{Hogg.et.al.2004.redSequenceWithOverdensity}.  The slope stays
constant throughout our analysis, $i.e.$ the slope does not depend on
galaxy selection method nor redshift bin.  However, the galaxies with
EW(H$\alpha$)$<0$\AA\ have the narrowest and most symmetric
distributions in color and D4000 index, and their stacked spectra show
no sign of emission lines.  Also, all of the H$\alpha$ selected
galaxies have S\'ersic $n\geqslant2$, $i.e.$ they are all
bulge-dominated systems.

In comparison, the early-type (S\'ersic $n>2$) samples include a
significant number of blue/star-forming galaxies and have a color
scatter that is twice as large as in the H$\alpha$ sample, $e.g.$
observed $\sigma_{\tiny{\textnormal{MAD}}}(g'-r')=0.077$ vs. 0.038 at
$z\sim0.15$.  Their stacked spectra also show strong emission lines.
Using D4000 to estimate mean stellar age, we find that the average
$n>2$ galaxy is only two-thirds as old as the average
H$\alpha$-selected galaxy, $e.g.$ 3.0 vs. 4.5 Gyr at $z\sim0.15$.

Applying a color-cut to the morphologically-selected sample does
reduce the fraction of star-forming galaxies.  However, the color and
D4000 distributions are still wider and more asymmetric than in the
H$\alpha$ sample, $i.e.$ the sample still contains active
galaxies. From the spectral line ratios, we find that the early-type
galaxies with emission lines tend to have AGN rather than ongoing star
formation.

Our analysis confirms that selecting galaxies with
EW(H$\alpha$)$<0$\AA\ is the most effective method for identifying an
extremely pure, homogeneously aged galaxy population.  Note that with
the development of near-infrared spectrographs, H$\alpha$ can be
measured to $z\sim2$ whereas an imaging resolution of $<0.1''$ is
needed to quantify galaxy morphology at $z>0.5$.  By isolating the
truly quiescent galaxies that populate and stay on the red sequence as
a function of redshift, we are better able to quantify how their
number density and masses evolve.

\acknowledgments

K. Tran thanks Marijn Franx for the thoughtful discussions that lead
to this project.  The authors acknowledge support from the Swiss
National Science Foundation (grant PP002-110576).


\bibliographystyle{apj}
\bibliography{PaperWVLA}

\clearpage

\clearpage

\begin{deluxetable}{rrr}
\tabletypesize{\scriptsize}
\tablewidth{0pt}
\tablecaption{Main Galaxy (MG) Sample}
\tablehead{\colhead{$\vartriangle$\textit{z}}      & \colhead{$N_{gal}$} & 	\colhead{$M_{r'}$\tablenotemark{a}}
}
\startdata
 0.050 $<z\leq$ 0.055 & 13,167 & -18.5 \\
 0.055 $<z\leq$ 0.060 & 12,283 & -18.7 \\
 0.060 $<z\leq$ 0.065 & 15,959 & -18.9 \\
 0.065 $<z\leq$ 0.070 & 16,207 & -19.1 \\
 0.070 $<z\leq$ 0.075 & 18,820 & -19.3 \\
 0.075 $<z\leq$ 0.080 & 19,408 & -19.4 \\
 0.080 $<z\leq$ 0.085 & 19,675 & -19.6 \\
 0.085 $<z\leq$ 0.090 & 16,721 & -19.7 \\
 0.090 $<z\leq$ 0.095 & 14,296 & -19.8 \\
 0.095 $<z\leq$ 0.100 & 14,094 & -20.0 \\
 0.100 $<z\leq$ 0.105 & 13,324 & -20.1 \\
 0.105 $<z\leq$ 0.110 & 13,210 & -20.2 \\
 0.110 $<z\leq$ 0.115 & 14,172 & -20.3 \\
 0.115 $<z\leq$ 0.120 & 13,193 & -20.4 \\
 0.120 $<z\leq$ 0.125 & 11,580 & -20.5 \\
 0.125 $<z\leq$ 0.130 & 11,807 & -20.6 \\
 0.130 $<z\leq$ 0.135 & 11,974 & -20.7 \\
 0.135 $<z\leq$ 0.140 & 10,736 & -20.8 \\
 0.140 $<z\leq$ 0.145 &  9,264 & -20.9 \\
 0.145 $<z\leq$ 0.150 &  8,641 & -21.0 \\
\enddata
\tablenotetext{a}{The applied $r'$-band absolute Petrosian magnitude
cuts of each redshift bin. This absolute magnitude includes a
K-correction for a ``typical'' H$\alpha$ passive galaxy (early-type
SED). These values are drawn as short solid horizontal lines in Figure
\ref{fig.Absolute magnitudes cut-off}.}
\label{tbl.Main galaxy sample}
\end{deluxetable}

\begin{deluxetable}{lcccc}
\tabletypesize{\scriptsize}
\tablewidth{0pt}
\tablecaption{Selection Criteria for Main Galaxy (MG) Sample\tablenotemark{a}}
\tablehead{
\colhead{Sample} & \colhead{Selection} & \colhead{EW H$\alpha$} & \colhead{S\'{e}rsic $n$} & \colhead{color-cut} \\
\colhead{} & \colhead{} & \colhead{[\AA{}]} & \colhead{} & \colhead{} 
}
\startdata
MG1 & H$\alpha$ & $<$ 0 & \nodata & \nodata \\
MG2 & S\'{e}rsic & \nodata & $>$ 2 & \nodata \\
MG3 & S\'{e}rsic + color-cut & \nodata & $>$ 2 & 3$\sigma_{\tiny{\textnormal{OBS}}}$ of CMR \\
\enddata
\tablenotetext{a}{We consider only galaxies that have signal-to-noise
ratios $\langle$S/N$\rangle_{g'} >$ 3, 
$\langle$S/N$\rangle_{r'} >$ 3, $\langle$S/N$\rangle_{spec} >$ 5, and
(S/N)$_{\tiny{\textnormal{H}\alpha}} >$ 2. The
3$\sigma_{\tiny{\textnormal{OBS}}}$ color-cut is based on MG1's
color-magnitude relation (CMR) and its color deviations
$\vartriangle$($g'-r'$). MG1, MG2, and MG3 have an increasing
magnitude limit in $M_{r'}$ with increasing redshift
(redshift bins of $\vartriangle$\textit{z} = 0.005; see Table
\ref{tbl.Main galaxy sample} \& Figure \ref{fig.Absolute magnitudes
cut-off}).}   
\label{tbl.Criteria}
\end{deluxetable}

\begin{deluxetable}{lccccc}
\tabletypesize{\scriptsize}
\tablewidth{0pt}
\tablecaption{Selection Criteria for Luminous Galaxy (LG) Sample\tablenotemark{a}}
\tablehead{
\colhead{Sample} & \colhead{Selection} & \colhead{EW H$\alpha$} &
\colhead{S\'{e}rsic $n$} & \colhead{color-cut} & \colhead{$N_{gal}$} \\
\colhead{} & \colhead{} & \colhead{[\AA{}]} & \colhead{} & \colhead{}
& \colhead{} }
\startdata
LG1 & H$\alpha$ & $<$ 0 & \nodata & \nodata & 30,423\\
LG2 & S\'{e}rsic & \nodata & $>$ 2 & \nodata & 67,729\\
LG3 & S\'{e}rsic + color-cut & \nodata & $>$ 2 & 3$\sigma_{\tiny{\textnormal{OBS}}}$ of CMR & 46,266\\
\enddata
\tablenotetext{a}{The Luminous Galaxy (LG) sample includes 81,232 objects. LG1, LG2 \& LG3 apply the same selection criteria as the MG samples listed in Table \ref{tbl.Criteria} with an additional
cut in absolute magnitude ($M_{r'} < -21.0$) used over the
entire redshift range of $0.05<z\leq0.15$. }
\label{tbl.Luminous galaxy sample}
\end{deluxetable}

\begin{deluxetable}{rrrrrr}
\tabletypesize{\scriptsize}
\tablewidth{0pt}
\tablecaption{Color-magnitude relation: MG1}
\tablehead{
\colhead{$\vartriangle$z} & \colhead{f ($N_{gal}$)\tablenotemark{a}} & \colhead{$m$ $\pm \sigma_{m}$\tablenotemark{b}} & \colhead{$n$ $\pm \sigma_{n}$\tablenotemark{b}} & \colhead{$\sigma_{\tiny{\textnormal{OBS}}}$($g'-r'$)\tablenotemark{c}} & \colhead{$\sigma_{\tiny{\textnormal{INT}}}$($g'-r'$)\tablenotemark{d}}}
\startdata
   0.050 $<z\leq$ 0.055 & 22.5\% (2,963) &  -0.017 $\pm$0.001 & 1.12 $\pm$0.02 & 0.0295 $\pm$0.0007 & 0.0287 $\pm$0.0007 \\
   0.055 $<z\leq$ 0.060 & 23.4\% (2,879) &  -0.016 $\pm$0.002 & 1.12 $\pm$0.03 & 0.0298 $\pm$0.0008 & 0.0289 $\pm$0.0008 \\
   0.060 $<z\leq$ 0.065 & 23.8\% (3,801) &  -0.017 $\pm$0.001 & 1.15 $\pm$0.02 & 0.0292 $\pm$0.0006 & 0.0283 $\pm$0.0006 \\
   0.065 $<z\leq$ 0.070 & 25.4\% (4,113) &  -0.018 $\pm$0.001 & 1.17 $\pm$0.02 & 0.0292 $\pm$0.0006 & 0.0282 $\pm$0.0006 \\
   0.070 $<z\leq$ 0.075 & 27.5\% (5,179) &  -0.017 $\pm$0.001 & 1.17 $\pm$0.02 & 0.0296 $\pm$0.0005 & 0.0286 $\pm$0.0005 \\
   0.075 $<z\leq$ 0.080 & 28.0\% (5,433) &  -0.018 $\pm$0.001 & 1.20 $\pm$0.02 & 0.0292 $\pm$0.0005 & 0.0281 $\pm$0.0005 \\
   0.080 $<z\leq$ 0.085 & 29.5\% (5,799) &  -0.016 $\pm$0.001 & 1.18 $\pm$0.02 & 0.0305 $\pm$0.0005 & 0.0293 $\pm$0.0005 \\
   0.085 $<z\leq$ 0.090 & 29.5\% (4,940) &  -0.015 $\pm$0.002 & 1.17 $\pm$0.03 & 0.0315 $\pm$0.0005 & 0.0302 $\pm$0.0005 \\
   0.090 $<z\leq$ 0.095 & 30.0\% (4,286) &  -0.016 $\pm$0.002 & 1.21 $\pm$0.03 & 0.0323 $\pm$0.0006 & 0.0311 $\pm$0.0006 \\
   0.095 $<z\leq$ 0.100 & 30.0\% (4,231) &  -0.016 $\pm$0.002 & 1.22 $\pm$0.04 & 0.0328 $\pm$0.0007 & 0.0315 $\pm$0.0007 \\
   0.100 $<z\leq$ 0.105 & 27.8\% (3,708) &  -0.018 $\pm$0.002 & 1.27 $\pm$0.03 & 0.0328 $\pm$0.0007 & 0.0314 $\pm$0.0006 \\
   0.105 $<z\leq$ 0.110 & 30.0\% (3,964) &  -0.022 $\pm$0.002 & 1.34 $\pm$0.04 & 0.0338 $\pm$0.0007 & 0.0324 $\pm$0.0006 \\
   0.110 $<z\leq$ 0.115 & 31.4\% (4,455) &  -0.022 $\pm$0.003 & 1.36 $\pm$0.05 & 0.0326 $\pm$0.0006 & 0.0311 $\pm$0.0006 \\
   0.115 $<z\leq$ 0.120 & 28.8\% (3,803) &  -0.015 $\pm$0.003 & 1.26 $\pm$0.04 & 0.0362 $\pm$0.0007 & 0.0347 $\pm$0.0007 \\
   0.120 $<z\leq$ 0.125 & 28.7\% (3,320) &  -0.020 $\pm$0.003 & 1.37 $\pm$0.05 & 0.0359 $\pm$0.0009 & 0.0343 $\pm$0.0009 \\
   0.125 $<z\leq$ 0.130 & 31.6\% (3,729) &  -0.024 $\pm$0.003 & 1.44 $\pm$0.05 & 0.0368 $\pm$0.0007 & 0.0352 $\pm$0.0007 \\
   0.130 $<z\leq$ 0.135 & 34.5\% (4,129) &  -0.021 $\pm$0.003 & 1.41 $\pm$0.05 & 0.0375 $\pm$0.0008 & 0.0358 $\pm$0.0008 \\
   0.135 $<z\leq$ 0.140 & 36.2\% (3,890) &  -0.024 $\pm$0.003 & 1.47 $\pm$0.05 & 0.0365 $\pm$0.0008 & 0.0346 $\pm$0.0007 \\
   0.140 $<z\leq$ 0.145 & 36.8\% (3,413) &  -0.023 $\pm$0.003 & 1.48 $\pm$0.06 & 0.0369 $\pm$0.0009 & 0.0349 $\pm$0.0008 \\
   0.145 $<z\leq$ 0.150 & 36.7\% (3,171) &  -0.028 $\pm$0.004 & 1.58 $\pm$0.06 & 0.0377 $\pm$0.0009 & 0.0357 $\pm$0.0009 \\

\enddata
\tablenotetext{a}{The fraction of EW H$\alpha$ $<$ 0\AA{} galaxies
inside each redshift bin. The value in parentheses is the selected
number of galaxies (see Table \ref{tbl.Main galaxy sample}).}
\tablenotetext{b}{The slope ($m$) and normalization ($n$) of the
linear color-magnitude relation ($g'-r'$) = $n + mr'$ and its 1$\sigma$
errors; we fit the CM relation using apparent magnitudes for all of
the Main Galaxy samples (see \S\ref{scatter} for details).} 
\tablenotetext{c}{The observed scatter in color and its
error; we use the Median Absolute Deviation (MAD) to characterize the
color scatter and calculate its error using the bootstrap resampling
method (\citealt{Efron.1979.Bootstrap}).}  
\tablenotetext{d}{The intrinsic scatter in color and 
its error (see \S\ref{scatter} for details). }
\label{tbl.CMrelation slopes Ha<0}
\end{deluxetable}

\begin{deluxetable}{rrrrrr}
\tabletypesize{\scriptsize}
\tablewidth{0pt}
\tablecaption{Color-magnitude relation: MG2}
\tablehead{
\colhead{$\vartriangle$z} & \colhead{f ($N_{gal}$)\tablenotemark{a}} & \colhead{m $\pm \sigma_{m}$\tablenotemark{b}} & \colhead{n $\pm \sigma_{n}$\tablenotemark{b}} & \colhead{$\sigma_{\tiny{\textnormal{OBS}}}$($g'-r'$)\tablenotemark{c}} & \colhead{$\sigma_{\tiny{\textnormal{INT}}}$($g'-r'$)\tablenotemark{d}}}
\startdata
   0.050 $<z\leq$ 0.055 & 50.9\% ( 6,703) & -0.017 $\pm$0.002 & 1.12 $\pm$0.04 & 0.0686 $\pm$0.0014 & 0.0682 $\pm$0.0014 \\
   0.055 $<z\leq$ 0.060 & 52.2\% ( 6,410) & -0.017 $\pm$0.002 & 1.12 $\pm$0.04 & 0.0661 $\pm$0.0015 & 0.0657 $\pm$0.0015 \\
   0.060 $<z\leq$ 0.065 & 54.7\% ( 8,730) & -0.020 $\pm$0.002 & 1.18 $\pm$0.04 & 0.0646 $\pm$0.0012 & 0.0642 $\pm$0.0012 \\
   0.065 $<z\leq$ 0.070 & 55.6\% ( 9,019) & -0.019 $\pm$0.002 & 1.19 $\pm$0.04 & 0.0640 $\pm$0.0010 & 0.0636 $\pm$0.0010 \\
   0.070 $<z\leq$ 0.075 & 58.9\% (11,081) & -0.017 $\pm$0.002 & 1.17 $\pm$0.03 & 0.0604 $\pm$0.0010 & 0.0599 $\pm$0.0010 \\
   0.075 $<z\leq$ 0.080 & 60.5\% (11,735) & -0.017 $\pm$0.002 & 1.18 $\pm$0.03 & 0.0604 $\pm$0.0009 & 0.0598 $\pm$0.0008 \\
   0.080 $<z\leq$ 0.085 & 63.0\% (12,395) & -0.018 $\pm$0.002 & 1.21 $\pm$0.04 & 0.0622 $\pm$0.0009 & 0.0616 $\pm$0.0009 \\
   0.085 $<z\leq$ 0.090 & 63.7\% (10,652) & -0.017 $\pm$0.002 & 1.20 $\pm$0.04 & 0.0633 $\pm$0.0010 & 0.0627 $\pm$0.0010 \\
   0.090 $<z\leq$ 0.095 & 63.8\% ( 9,119) & -0.014 $\pm$0.003 & 1.17 $\pm$0.05 & 0.0663 $\pm$0.0011 & 0.0657 $\pm$0.0011 \\
   0.095 $<z\leq$ 0.100 & 65.8\% ( 9,267) & -0.018 $\pm$0.003 & 1.25 $\pm$0.05 & 0.0660 $\pm$0.0012 & 0.0653 $\pm$0.0012 \\
   0.100 $<z\leq$ 0.105 & 65.7\% ( 8,755) & -0.022 $\pm$0.003 & 1.31 $\pm$0.05 & 0.0682 $\pm$0.0012 & 0.0675 $\pm$0.0012 \\
   0.105 $<z\leq$ 0.110 & 67.3\% ( 8,887) & -0.023 $\pm$0.003 & 1.36 $\pm$0.05 & 0.0680 $\pm$0.0012 & 0.0673 $\pm$0.0012 \\
   0.110 $<z\leq$ 0.115 & 69.9\% ( 9,906) & -0.025 $\pm$0.003 & 1.39 $\pm$0.05 & 0.0700 $\pm$0.0011 & 0.0692 $\pm$0.0011 \\
   0.115 $<z\leq$ 0.120 & 70.5\% ( 9,300) & -0.024 $\pm$0.003 & 1.39 $\pm$0.06 & 0.0738 $\pm$0.0011 & 0.0731 $\pm$0.0011 \\
   0.120 $<z\leq$ 0.125 & 70.2\% ( 8,124) & -0.027 $\pm$0.004 & 1.46 $\pm$0.06 & 0.0765 $\pm$0.0014 & 0.0757 $\pm$0.0014 \\
   0.125 $<z\leq$ 0.130 & 72.4\% ( 8,545) & -0.029 $\pm$0.004 & 1.51 $\pm$0.06 & 0.0766 $\pm$0.0013 & 0.0758 $\pm$0.0013 \\
   0.130 $<z\leq$ 0.135 & 75.3\% ( 9,013) & -0.027 $\pm$0.004 & 1.50 $\pm$0.06 & 0.0763 $\pm$0.0014 & 0.0755 $\pm$0.0014 \\
   0.135 $<z\leq$ 0.140 & 76.7\% ( 8,237) & -0.029 $\pm$0.004 & 1.55 $\pm$0.07 & 0.0762 $\pm$0.0014 & 0.0753 $\pm$0.0014 \\
   0.140 $<z\leq$ 0.145 & 77.4\% ( 7,169) & -0.031 $\pm$0.005 & 1.60 $\pm$0.08 & 0.0761 $\pm$0.0014 & 0.0752 $\pm$0.0014 \\
   0.145 $<z\leq$ 0.150 & 78.5\% ( 6,781) & -0.039 $\pm$0.005 & 1.75 $\pm$0.09 & 0.0771 $\pm$0.0016 & 0.0761 $\pm$0.0016 \\

\enddata
\tablenotetext{a}{The fraction of S\'{e}rsic $n >$ 2 galaxies inside
each redshift bin. The value in parentheses is the selected number of
galaxies (see Table \ref{tbl.Main galaxy sample}).}
\tablenotetext{b}{The slope ($m$) and normalization ($n$) of the
linear color-magnitude relation ($g'-r'$) = $n + mr'$ and its 1$\sigma$
errors; we fit the CM relation using apparent magnitudes for all of
the Main Galaxy samples (see \S\ref{scatter} for details).} 
\tablenotetext{c}{The observed scatter in color and its
error; we use the Median Absolute Deviation (MAD) to characterize the
color scatter and calculate its error using the bootstrap resampling
method (\citealt{Efron.1979.Bootstrap}).}  
\tablenotetext{d}{The intrinsic scatter in color and 
its error (see \S\ref{scatter} for details). }
\label{tbl.CMrelation slopes Sersic}
\end{deluxetable}

\begin{deluxetable}{rrrrrr}
\tabletypesize{\scriptsize}
\tablewidth{0pt}
\tablecaption{Color-magnitude relation: MG3}
\tablehead{
\colhead{$\vartriangle$z} & \colhead{f ($N_{gal}$)\tablenotemark{a}} & \colhead{m $\pm \sigma_{m}$\tablenotemark{b}} & \colhead{n $\pm \sigma_{n}$\tablenotemark{b}} & \colhead{$\sigma_{\tiny{\textnormal{OBS}}}$($g'-r'$)\tablenotemark{c}} & \colhead{$\sigma_{\tiny{\textnormal{INT}}}$($g'-r'$)\tablenotemark{d}}}
\startdata
   0.050 $<z\leq$ 0.055 & 32.3\% (4,252) & -0.018 $\pm$0.001 & 1.13 $\pm$0.02 & 0.0321 $\pm$0.0007 & 0.0313 $\pm$0.0007 \\
   0.055 $<z\leq$ 0.060 & 33.8\% (4,153) & -0.017 $\pm$0.001 & 1.12 $\pm$0.02 & 0.0323 $\pm$0.0006 & 0.0314 $\pm$0.0006 \\
   0.060 $<z\leq$ 0.065 & 35.5\% (5,671) & -0.020 $\pm$0.001 & 1.18 $\pm$0.02 & 0.0319 $\pm$0.0005 & 0.0310 $\pm$0.0005 \\
   0.065 $<z\leq$ 0.070 & 36.0\% (5,831) & -0.019 $\pm$0.001 & 1.18 $\pm$0.02 & 0.0312 $\pm$0.0005 & 0.0302 $\pm$0.0005 \\
   0.070 $<z\leq$ 0.075 & 39.4\% (7,407) & -0.017 $\pm$0.001 & 1.16 $\pm$0.02 & 0.0313 $\pm$0.0005 & 0.0302 $\pm$0.0005 \\
   0.075 $<z\leq$ 0.080 & 39.7\% (7,710) & -0.017 $\pm$0.001 & 1.18 $\pm$0.02 & 0.0301 $\pm$0.0004 & 0.0289 $\pm$0.0004 \\
   0.080 $<z\leq$ 0.085 & 42.4\% (8,334) & -0.017 $\pm$0.001 & 1.20 $\pm$0.02 & 0.0324 $\pm$0.0004 & 0.0312 $\pm$0.0004 \\
   0.085 $<z\leq$ 0.090 & 43.0\% (7,183) & -0.016 $\pm$0.001 & 1.19 $\pm$0.02 & 0.0332 $\pm$0.0005 & 0.0319 $\pm$0.0005 \\
   0.090 $<z\leq$ 0.095 & 42.5\% (6,071) & -0.016 $\pm$0.002 & 1.20 $\pm$0.03 & 0.0342 $\pm$0.0005 & 0.0329 $\pm$0.0005 \\
   0.095 $<z\leq$ 0.100 & 44.4\% (6,261) & -0.019 $\pm$0.001 & 1.26 $\pm$0.03 & 0.0353 $\pm$0.0005 & 0.0339 $\pm$0.0005 \\
   0.100 $<z\leq$ 0.105 & 43.5\% (5,796) & -0.020 $\pm$0.002 & 1.30 $\pm$0.03 & 0.0351 $\pm$0.0005 & 0.0337 $\pm$0.0005 \\
   0.105 $<z\leq$ 0.110 & 45.3\% (5,980) & -0.024 $\pm$0.002 & 1.38 $\pm$0.03 & 0.0364 $\pm$0.0006 & 0.0350 $\pm$0.0005 \\
   0.110 $<z\leq$ 0.115 & 45.8\% (6,485) & -0.023 $\pm$0.002 & 1.38 $\pm$0.04 & 0.0346 $\pm$0.0005 & 0.0330 $\pm$0.0005 \\
   0.115 $<z\leq$ 0.120 & 46.7\% (6,167) & -0.019 $\pm$0.002 & 1.32 $\pm$0.03 & 0.0392 $\pm$0.0006 & 0.0377 $\pm$0.0006 \\
   0.120 $<z\leq$ 0.125 & 46.0\% (5,322) & -0.024 $\pm$0.002 & 1.41 $\pm$0.04 & 0.0396 $\pm$0.0007 & 0.0380 $\pm$0.0007 \\
   0.125 $<z\leq$ 0.130 & 47.9\% (5,652) & -0.026 $\pm$0.002 & 1.47 $\pm$0.04 & 0.0391 $\pm$0.0006 & 0.0374 $\pm$0.0006 \\
   0.130 $<z\leq$ 0.135 & 50.4\% (6,035) & -0.026 $\pm$0.002 & 1.48 $\pm$0.04 & 0.0401 $\pm$0.0007 & 0.0383 $\pm$0.0007 \\
   0.135 $<z\leq$ 0.140 & 50.6\% (5,430) & -0.026 $\pm$0.002 & 1.51 $\pm$0.04 & 0.0387 $\pm$0.0007 & 0.0367 $\pm$0.0006 \\
   0.140 $<z\leq$ 0.145 & 51.7\% (4,789) & -0.028 $\pm$0.003 & 1.54 $\pm$0.05 & 0.0398 $\pm$0.0007 & 0.0378 $\pm$0.0007 \\
   0.145 $<z\leq$ 0.150 & 52.2\% (4,509) & -0.035 $\pm$0.003 & 1.69 $\pm$0.05 & 0.0400 $\pm$0.0008 & 0.0379 $\pm$0.0008 \\

\enddata
\tablenotetext{a}{The fraction of galaxies that have S\'{e}rsic $n>2$
and color deviation $\vartriangle(g'-r') > -3\sigma_{\tiny{\textnormal{OBS}}}$ in each redshift 
bin (see Figure \ref{fig.CMrelation slopes Ha+Sersic}, middle and
bottom panels). The value in parentheses is the selected number of
galaxies (see Table \ref{tbl.Criteria}).}
\tablenotetext{b}{The slope ($m$) and normalization ($n$) of the
linear color-magnitude relation ($g'-r'$) = $n + mr'$ and its 1$\sigma$
errors; we fit the CM relation using apparent magnitudes for all of
the Main Galaxy samples (see \S\ref{scatter} for details).} 
\tablenotetext{c}{The observed scatter in color and its
error; we use the Median Absolute Deviation (MAD) to characterize the
color scatter and calculate its error using the bootstrap resampling
method (\citealt{Efron.1979.Bootstrap}).}  
\tablenotetext{d}{The intrinsic scatter in color and 
its error (see \S\ref{scatter} for details). }
\label{tbl.CMrelation slopes Sersic+color-cut}
\end{deluxetable}

\begin{deluxetable}{rrrrr}
\tabletypesize{\scriptsize}
\tablewidth{0pt}
\tablecaption{Active Fractions: MG2}
\tablehead{
\colhead{$\vartriangle$z} & \colhead{f ($N_{\tiny{\textnormal{active}}}$)\tablenotemark{a}} & \colhead{f ($N_{\tiny{\textnormal{AGN}}}$)\tablenotemark{b}} & \colhead{f ($N_{\tiny{\textnormal{AGN/SF}}}$)\tablenotemark{c}} & \colhead{f ($N_{\tiny{\textnormal{SF}}}$)\tablenotemark{d}}}
\startdata
 0.050$<z\leq$0.055 & 31.5\% (2,114) & 5.8\% (390) &  11.1\% (  747) & 14.6\% (977) \\
 0.055$<z\leq$0.060 & 29.6\% (1,896) & 5.7\% (363) &  11.4\% (  732) & 12.5\% (801) \\
 0.060$<z\leq$0.065 & 28.4\% (2,477) & 5.9\% (518) &  11.4\% (  991) & 11.1\% (968) \\
 0.065$<z\leq$0.070 & 26.8\% (2,421) & 5.3\% (479) &  11.2\% (1,007) & 10.4\% (935) \\
 0.070$<z\leq$0.075 & 24.4\% (2,703) & 5.1\% (566) &  10.3\% (1,139) &  9.0\% (998) \\
 0.075$<z\leq$0.080 & 22.9\% (2,683) & 4.9\% (574) &  10.2\% (1,195) &  7.8\% (914) \\
 0.080$<z\leq$0.085 & 21.9\% (2,720) & 5.0\% (622) &  10.3\% (1,275) &  6.6\% (823) \\
 0.085$<z\leq$0.090 & 20.7\% (2,202) & 4.4\% (472) &  10.1\% (1,081) &  6.1\% (649) \\
 0.090$<z\leq$0.095 & 20.4\% (1,857) & 4.9\% (443) &   9.7\% (  886) &  5.8\% (528) \\
 0.095$<z\leq$0.100 & 20.7\% (1,914) & 5.8\% (536) &   9.8\% (  908) &  5.1\% (470) \\
 0.100$<z\leq$0.105 & 20.5\% (1,798) & 5.8\% (506) &   9.7\% (  850) &  5.0\% (442) \\
 0.105$<z\leq$0.110 & 19.1\% (1,693) & 5.3\% (468) &   9.0\% (  800) &  4.8\% (425) \\
 0.110$<z\leq$0.115 & 16.0\% (1,586) & 5.4\% (538) &   7.1\% (  706) &  3.5\% (342) \\
 0.115$<z\leq$0.120 & 17.6\% (1,633) & 5.1\% (478) &   8.4\% (  784) &  4.0\% (371) \\
 0.120$<z\leq$0.125 & 17.6\% (1,431) & 5.2\% (423) &   8.2\% (  665) &  4.2\% (343) \\
 0.125$<z\leq$0.130 & 17.1\% (1,462) & 5.1\% (435) &   7.9\% (  672) &  4.2\% (355) \\
 0.130$<z\leq$0.135 & 14.9\% (1,343) & 4.8\% (432) &   7.0\% (  628) &  3.1\% (283) \\
 0.135$<z\leq$0.140 & 14.5\% (1,191) & 4.5\% (369) &   7.1\% (  586) &  2.9\% (236) \\
 0.140$<z\leq$0.145 & 12.9\% (  926) & 4.5\% (323) &   6.1\% (  438) &  2.3\% (165) \\
 0.145$<z\leq$0.150 & 12.6\% (  853) & 4.3\% (292) &   6.2\% (  418) &  2.1\% (143) \\

\enddata
\tablecomments{MG2 only consists of galaxies that have S\'{e}rsic n
$>$ 2 in the $r'$-band. The fractions are relative to the total number of
MG2 galaxies in the same redshift bin.} 
\tablenotetext{a}{The fraction of galaxies inside each redshift bin of
the MG sample that have emission in H$\alpha$, H$\beta$,
[NII], and [OIII] with a S/N $>$ 2. The value in parentheses is the
selected number of galaxies.}
\tablenotetext{b}{The fraction of pure AGN dominated galaxies 
identified by the separation line of \cite{Kewley.et.al.2001.BPTplot}
(see Fig.~\ref{fig.BPTplot Sersic+Sersic-cut}).} 
\tablenotetext{c}{The fraction of mixed AGN/starforming galaxies
identified by the separation lines of
\cite{Kauffmann.et.al.2003.BPTplot} and
\cite{Kewley.et.al.2001.BPTplot} (see Fig. \ref{fig.BPTplot
Sersic+Sersic-cut}).} 
\tablenotetext{d}{The fraction of star-forming galaxies identified by the separation line of \cite{Kauffmann.et.al.2003.BPTplot} (see Fig. \ref{fig.BPTplot Sersic+Sersic-cut}). }
\label{tbl.BPT Sersic}
\end{deluxetable}

\begin{deluxetable}{rrrrr}
\tabletypesize{\scriptsize}
\tablewidth{0pt}
\tablecaption{Active Fractions: MG3}
\tablehead{
\colhead{$\vartriangle$z} & \colhead{f ($N_{\tiny{\textnormal{active}}}$)\tablenotemark{a}} & \colhead{f ($N_{\tiny{\textnormal{AGN}}}$)\tablenotemark{b}} & \colhead{f ($N_{\tiny{\textnormal{AGN/SF}}}$)\tablenotemark{c}} & \colhead{f ($N_{\tiny{\textnormal{SF}}}$)\tablenotemark{d}}}
\startdata
 0.050$<z\leq$0.055 & 9.3\% (394) & 3.6\% (152) & 4.4\% (186) & 1.3\% (56) \\
 0.055$<z\leq$0.060 & 9.0\% (372) & 3.5\% (144) & 4.4\% (181) & 1.1\% (47) \\
 0.060$<z\leq$0.065 & 8.4\% (474) & 3.6\% (202) & 3.8\% (216) & 1.0\% (56) \\
 0.065$<z\leq$0.070 & 7.4\% (431) & 3.1\% (183) & 3.6\% (208) & 0.7\% (40) \\
 0.070$<z\leq$0.075 & 6.2\% (460) & 2.5\% (184) & 3.1\% (229) & 0.6\% (47) \\
 0.075$<z\leq$0.080 & 5.4\% (416) & 2.2\% (169) & 2.8\% (216) & 0.4\% (31) \\
 0.080$<z\leq$0.085 & 5.5\% (460) & 2.2\% (186) & 3.0\% (246) & 0.3\% (28) \\
 0.085$<z\leq$0.090 & 5.0\% (362) & 2.0\% (147) & 2.7\% (195) & 0.3\% (20) \\
 0.090$<z\leq$0.095 & 4.8\% (291) & 2.2\% (135) & 2.2\% (131) & 0.4\% (25) \\
 0.095$<z\leq$0.100 & 4.8\% (299) & 2.3\% (145) & 2.2\% (136) & 0.3\% (18) \\
 0.100$<z\leq$0.105 & 4.5\% (262) & 2.5\% (145) & 1.8\% (105) & 0.2\% (12) \\
 0.105$<z\leq$0.110 & 4.1\% (243) & 2.2\% (130) & 1.7\% (101) & 0.2\% (12) \\
 0.110$<z\leq$0.115 & 2.9\% (189) & 1.9\% (125) & 0.9\% ( 58) & 0.1\% ( 6) \\
 0.115$<z\leq$0.120 & 3.6\% (221) & 2.2\% (138) & 1.2\% ( 72) & 0.2\% (11) \\
 0.120$<z\leq$0.125 & 3.8\% (204) & 2.0\% (106) & 1.7\% ( 90) & 0.2\% ( 8) \\
 0.125$<z\leq$0.130 & 3.1\% (178) & 1.7\% ( 96) & 1.2\% ( 70) & 0.2\% (12) \\
 0.130$<z\leq$0.135 & 2.6\% (157) & 1.6\% ( 99) & 0.8\% ( 50) & 0.1\% ( 8) \\
 0.135$<z\leq$0.140 & 2.2\% (118) & 1.5\% ( 81) & 0.7\% ( 36) & 0.0\% ( 1) \\
 0.140$<z\leq$0.145 & 1.8\% ( 88) & 1.4\% ( 66) & 0.4\% ( 20) & 0.0\% ( 2) \\
 0.145$<z\leq$0.150 & 2.7\% (121) & 1.8\% ( 81) & 0.8\% ( 36) & 0.1\% ( 4) \\

\enddata
\tablecomments{MG3 only consists of galaxies that have S\'{e}rsic $n
>$ 2 in the $r'$-band and color deviation $\vartriangle(g'-r') >
-3\sigma_{\tiny{\textnormal{OBS}}}$ (see Figure \ref{fig.CMrelation
slopes Ha+Sersic}, middle and bottom panels). The fractions are
relative to the total number of MG3 galaxies in the same redshift
bin.} 
\tablenotetext{a}{The fraction of galaxies inside each redshift bin
that have emission in H$\alpha$, H$\beta$, [NII], and [OIII] with a
S/N $>$ 2. } 
\tablenotetext{b}{The fraction of pure AGN dominated galaxies
identified by the separation line of \cite{Kewley.et.al.2001.BPTplot}
(see Fig. \ref{fig.BPTplot Sersic+Sersic-cut}). } 
\tablenotetext{c}{The fraction of mixed AGN/starforming galaxies
identified by the separation lines of
\cite{Kauffmann.et.al.2003.BPTplot} and
\cite{Kewley.et.al.2001.BPTplot} (see Fig. \ref{fig.BPTplot
Sersic+Sersic-cut}).} 
\tablenotetext{d}{The fraction of star-forming galaxies identified by
the separation line of \cite{Kauffmann.et.al.2003.BPTplot} (see
Fig. \ref{fig.BPTplot Sersic+Sersic-cut}). } 
\label{tbl.BPT Sersic+color-cut}
\end{deluxetable}

\begin{deluxetable}{rrrrrrr}
\tabletypesize{\scriptsize}
\tablewidth{0pt}
\tablecaption{Ages from Median D4000: MG1, MG2, MG3}
\tablehead{
\colhead{$\vartriangle$\textit{z}} & \colhead{$\langle$D4000$\rangle_{\tiny{\textnormal{MG1}}}$\tablenotemark{a}} & \colhead{$\langle$Age$\rangle_{\tiny{\textnormal{MG1}}}$\tablenotemark{b}} & \colhead{$\langle$D4000$\rangle_{\tiny{\textnormal{MG2}}}$\tablenotemark{a}} & \colhead{$\langle$Age$\rangle_{\tiny{\textnormal{MG2}}}$\tablenotemark{b}} & \colhead{$\langle$D4000$\rangle_{\tiny{\textnormal{MG3}}}$\tablenotemark{a}} & \colhead{$\langle$Age$\rangle_{\tiny{\textnormal{MG3}}}$\tablenotemark{b}} \\
\colhead{} & \colhead{ } & \colhead{[Gyr]} & \colhead{} & \colhead{[Gyr]} & \colhead{} & \colhead{[Gyr]}
}
\startdata
 0.050$<z\leq$0.055 & 1.74 $\pm$0.07 & 3.25 $^{+1.25} _{-0.50}$ & 1.58 $\pm$0.16 & 2.00 $^{+1.25} _{-1.30}$ & 1.71 $\pm$0.09 & 3.00 $^{+1.50} _{-0.60}$ \\
\\
 0.055$<z\leq$0.060 & 1.75 $\pm$0.07 & 3.50 $^{+1.25} _{-0.75}$ & 1.60 $\pm$0.15 & 2.20 $^{+1.30} _{-1.40}$ & 1.71 $\pm$0.09 & 3.00 $^{+1.50} _{-0.50}$ \\
\\
 0.060$<z\leq$0.065 & 1.75 $\pm$0.07 & 3.50 $^{+1.25} _{-0.75}$ & 1.60 $\pm$0.14 & 2.20 $^{+1.30} _{-1.30}$ & 1.72 $\pm$0.08 & 3.00 $^{+1.50} _{-0.40}$ \\
\\
 0.065$<z\leq$0.070 & 1.75 $\pm$0.06 & 3.50 $^{+1.25} _{-0.75}$ & 1.61 $\pm$0.14 & 2.30 $^{+1.20} _{-1.40}$ & 1.72 $\pm$0.08 & 3.00 $^{+1.50} _{-0.40}$ \\
\\
 0.070$<z\leq$0.075 & 1.76 $\pm$0.06 & 3.75 $^{+1.00} _{-0.75}$ & 1.62 $\pm$0.13 & 2.40 $^{+1.10} _{-1.40}$ & 1.72 $\pm$0.08 & 3.00 $^{+1.50} _{-0.40}$ \\
\\
 0.075$<z\leq$0.080 & 1.76 $\pm$0.06 & 3.75 $^{+1.00} _{-0.75}$ & 1.63 $\pm$0.13 & 2.50 $^{+1.25} _{-1.25}$ & 1.73 $\pm$0.07 & 3.25 $^{+1.25} _{-0.50}$ \\
\\
 0.080$<z\leq$0.085 & 1.76 $\pm$0.06 & 3.75 $^{+1.00} _{-0.75}$ & 1.64 $\pm$0.12 & 2.60 $^{+1.15} _{-1.20}$ & 1.73 $\pm$0.07 & 3.25 $^{+1.25} _{-0.50}$ \\
\\
 0.085$<z\leq$0.090 & 1.77 $\pm$0.06 & 4.00 $^{+1.00} _{-1.00}$ & 1.64 $\pm$0.12 & 2.60 $^{+1.15} _{-1.50}$ & 1.74 $\pm$0.07 & 3.25 $^{+1.25} _{-0.50}$ \\
\\
 0.090$<z\leq$0.095 & 1.77 $\pm$0.06 & 4.00 $^{+1.00} _{-1.00}$ & 1.64 $\pm$0.12 & 2.60 $^{+1.15} _{-1.50}$ & 1.74 $\pm$0.07 & 3.25 $^{+1.25} _{-0.50}$ \\
\\
 0.095$<z\leq$0.100 & 1.77 $\pm$0.06 & 4.00 $^{+1.00} _{-1.00}$ & 1.64 $\pm$0.12 & 2.60 $^{+1.15} _{-1.50}$ & 1.74 $\pm$0.07 & 3.25 $^{+1.25} _{-0.50}$ \\
\\
 0.100$<z\leq$0.105 & 1.77 $\pm$0.06 & 4.00 $^{+1.00} _{-1.00}$ & 1.64 $\pm$0.13 & 2.60 $^{+1.40} _{-1.20}$ & 1.74 $\pm$0.07 & 3.25 $^{+1.25} _{-0.50}$ \\
\\
 0.105$<z\leq$0.110 & 1.77 $\pm$0.06 & 4.00 $^{+1.00} _{-1.00}$ & 1.64 $\pm$0.12 & 2.60 $^{+1.40} _{-1.50}$ & 1.74 $\pm$0.07 & 3.25 $^{+1.25} _{-0.50}$ \\
\\
 0.110$<z\leq$0.115 & 1.78 $\pm$0.06 & 4.00 $^{+1.25} _{-1.00}$ & 1.65 $\pm$0.12 & 2.75 $^{+1.25} _{-1.15}$ & 1.75 $\pm$0.07 & 3.50 $^{+1.25} _{-0.75}$ \\
\\
 0.115$<z\leq$0.120 & 1.78 $\pm$0.06 & 4.00 $^{+1.25} _{-1.00}$ & 1.65 $\pm$0.12 & 2.75 $^{+1.25} _{-1.15}$ & 1.75 $\pm$0.07 & 3.50 $^{+1.25} _{-0.75}$ \\
\\
 0.120$<z\leq$0.125 & 1.78 $\pm$0.06 & 4.00 $^{+1.25} _{-1.00}$ & 1.65 $\pm$0.12 & 2.75 $^{+1.25} _{-1.15}$ & 1.75 $\pm$0.07 & 3.50 $^{+1.25} _{-0.75}$ \\
\\
 0.125$<z\leq$0.130 & 1.78 $\pm$0.06 & 4.00 $^{+1.25} _{-1.00}$ & 1.66 $\pm$0.12 & 2.75 $^{+1.25} _{-1.10}$ & 1.76 $\pm$0.07 & 3.75 $^{+1.00} _{-0.75}$ \\
\\
 0.130$<z\leq$0.135 & 1.79 $\pm$0.06 & 4.25 $^{+1.00} _{-1.00}$ & 1.67 $\pm$0.11 & 2.75 $^{+1.25} _{-0.85}$ & 1.76 $\pm$0.06 & 3.75 $^{+1.00} _{-0.75}$ \\
\\
 0.135$<z\leq$0.140 & 1.79 $\pm$0.06 & 4.25 $^{+1.00} _{-1.00}$ & 1.68 $\pm$0.11 & 2.75 $^{+1.50} _{-0.85}$ & 1.77 $\pm$0.06 & 4.00 $^{+1.00} _{-1.00}$ \\
\\
 0.140$<z\leq$0.145 & 1.80 $\pm$0.06 & 4.50 $^{+1.00} _{-1.00}$ & 1.69 $\pm$0.11 & 3.00 $^{+1.25} _{-1.00}$ & 1.78 $\pm$0.06 & 4.00 $^{+1.25} _{-1.00}$ \\
\\
 0.145$<z\leq$0.150 & 1.80 $\pm$0.06 & 4.50 $^{+1.00} _{-1.00}$ & 1.69 $\pm$0.11 & 3.00 $^{+1.50} _{-1.00}$ & 1.79 $\pm$0.07 & 4.25 $^{+1.00} _{-1.25}$ \\

\enddata
\tablecomments{We use the stellar synthesis model code of
\cite{Bruzual.Charlot.2003.BC03} with solar metallicity (Z = 0.02),
Salpeter IMF, and
the Padova stellar evolution track to determine an age. }
\tablenotetext{a}{The median D4000-value for MG1, MG2, or MG3 (see
Fig. \ref{fig.D4000 plot of Ha<0}).} 
\tablenotetext{b}{The age derived from the sample's median
D4000-value.  The average age in the highest redshift bins is older
because we have applied an increasing luminosity cut-off for the MG
samples such that only the most luminous galaxies are included at
higher redshift; studies have shown that more luminous galaxies tend
to be older (\citealt{Gallazzi.et.al2006.ageMetalGalsLocalUniverseSDSS,
Graves.et.al.2007.RedSequence}).}

\label{tbl.averageD4000 values}
\end{deluxetable}

\begin{deluxetable}{lrrrrrr}
\tabletypesize{\scriptsize}
\tablewidth{0pt}
\tablecaption{Color-magnitude relations: LG1, LG2, LG3}
\tablehead{
\colhead{Galaxy sample} & \colhead{$\vartriangle$z} & \colhead{f ($N_{\textnormal{gal}}$)\tablenotemark{a}} & \colhead{m $\pm \sigma_{m}$\tablenotemark{b}} & \colhead{n $\pm \sigma_{n}$\tablenotemark{b}} & \colhead{$\sigma_{\tiny{\textnormal{OBS}}}$($g'-r'$)\tablenotemark{c}} & \colhead{$\sigma_{\tiny{\textnormal{INT}}}$($g'-r'$)\tablenotemark{d}}}
\startdata
LG1& 0.05$<z\leq$0.15 &  37.4\% (30,423) &  -0.019 $\pm$0.001 & 0.33 $\pm$0.02 & 0.0233 $\pm$1.7$\cdot10^{-4}$ & 0.0150 $\pm$1.4$\cdot10^{-4}$ \\
LG2 & 0.05$<z\leq$0.15 &  83.3\% (67,729) &  -0.025 $\pm$0.001 &  0.20 $\pm$0.02 & 0.0464 $\pm$2.8$\cdot10^{-4}$ & 0.0428 $\pm$2.7$\cdot10^{-4}$ \\
LG3 & 0.05$<z\leq$0.15 &  56.9\% (46.266) &  -0.022 $\pm$0.001 &  0.26 $\pm$0.02 & 0.0253 $\pm$1.5$\cdot10^{-4}$ & 0.0176 $\pm$1.2$\cdot10^{-4}$ \\
\enddata
\tablenotetext{a}{The galaxy fraction is calculated with respect to
the total galaxy number of the LG sample (81,323 galaxies). The value
in parentheses is the number of galaxies for a given selection (see
Table \ref{tbl.Luminous galaxy sample}).}
\tablenotetext{b}{The slope ($m$) and normalization ($n$) of the
linear color-magnitude relation ($g'-r'$) = $n + mr'$ and its 1$\sigma$
errors; we fit the CM relation using absolute Petrosian magnitudes for all of
the Luminous Galaxy samples (see \S\ref{scatter} for details).} 
\tablenotetext{c}{The observed scatter in color and its
error; we use the Median Absolute Deviation (MAD) to characterize the
color scatter and calculate its error using the bootstrap resampling
method (\citealt{Efron.1979.Bootstrap}).}  
\tablenotetext{d}{The intrinsic scatter in color and 
its error (see \S\ref{scatter} for details). }
\label{tbl.AbsMag CMrelations}
\end{deluxetable}

\end{document}